\documentclass[sigconf,screen]{acmart}

\AtBeginDocument{%
  }

\usepackage{balance}

\usepackage[normalem]{ulem}
\usepackage{algorithmic}
\usepackage{graphicx}
\usepackage{textcomp}
\usepackage{xcolor}
\usepackage{microtype}
\usepackage{braket}
\usepackage{soul}
\usepackage{booktabs}
\usepackage{adjustbox}

\usepackage{bm}
\usepackage{url}

\usepackage[linewidth=1pt]{mdframed}

\usepackage{tikz}
\newcommand*\circled[1]{\tikz[baseline=(char.base)]{
            \node[shape=circle,draw,inner sep=0.5pt] (char) {#1};}}
\usepackage{mathtools}

\usepackage{caption}
\usepackage{subcaption}
\usepackage{tabularx}

\acmSubmissionID{asplosb23main-p7-p}
\received{2022-07-07}
\received[accepted]{2022-09-22}
\copyrightyear{2023}
\acmYear{2023}
\setcopyright{acmcopyright}
\acmConference[ASPLOS '23]{Proceedings of the 28th ACM International Conference on Architectural Support for Programming Languages and Operating Systems, Volume 2}{March 25--29, 2023}{Vancouver, BC, Canada}
\acmBooktitle{Proceedings of the 28th ACM International Conference on Architectural Support for Programming Languages and Operating Systems, Volume 2 (ASPLOS '23), March 25--29, 2023, Vancouver, BC, Canada}
\acmPrice{15.00}
\acmDOI{10.1145/3575693.3575739}
\acmISBN{978-1-4503-9916-6/23/03}

\begin{document}

%%
%% The "title" command has an optional parameter,
%% allowing the author to define a "short title" to be used in page headers.
\title{Navigating the Dynamic Noise Landscape of Variational Quantum Algorithms with QISMET}

%%
%% The "author" command and its associated commands are used to define
%% the authors and their affiliations.
%% Of note is the shared affiliation of the first two authors, and the
%% "authornote" and "authornotemark" commands
%% used to denote shared contribution to the research.

\author{Gokul Subramanian Ravi}
\authornote{Correspondence: gravi@uchicago.edu}
\affiliation{%
  \institution{University of Chicago, USA}
    %\country{USA}
}
%\email{}

\author{Kaitlin Smith}
\affiliation{%
  \institution{University of Chicago, USA}
    %\country{USA}
}
%\email{@uchicago.edu}

\author{Jonathan M. Baker}
\affiliation{%
  \institution{University of Chicago, USA}
    %\country{USA}
}
%\email{@uchicago.edu}

\author{Tejas Kannan}
\affiliation{%
  \institution{University of Chicago, USA}
    %\country{USA}
}
%\email{@uchicago.edu}

\author{Nathan Earnest}
\affiliation{%
  \institution{IBM Quantum, USA}
    %\country{USA}
}
%\email{@uchicago.edu}

\author{Ali Javadi-Abhari}
\affiliation{%
  \institution{IBM Quantum, USA}
    %\country{USA}
}
%\email{@uchicago.edu}

\author{Henry Hoffmann}
\affiliation{%
  \institution{University of Chicago, USA}
    %\country{USA}
}
%\email{@uchicago.edu}

\author{Frederic T. Chong}
\affiliation{%
  \institution{University of Chicago, USA}
}
%\email{@uchicago.edu}

%%
%% By default, the full list of authors will be used in the page
%% headers. Often, this list is too long, and will overlap
%% other information printed in the page headers. This command allows
%% the author to define a more concise list
%% of authors' names for this purpose.
\renewcommand{\shortauthors}{G. Ravi, K. Smith, J.M. Baker, T. Kannan, N. Earnest, A. Javadi-Abhari, H. Hoffmann, F.T. Chong}

%%
%% The abstract is a short summary of the work to be presented in the
%% article.
\begin{abstract}
In the Noisy Intermediate Scale Quantum (NISQ) era, the dynamic nature of quantum systems causes noise sources to constantly vary over time.
Transient errors from the dynamic NISQ noise landscape are challenging to comprehend and are especially detrimental to classes of applications that are iterative and/or long-running, and therefore their timely mitigation is important for quantum advantage in real-world applications.

The most popular examples of iterative long-running quantum applications are variational quantum algorithms (VQAs).
Iteratively, VQA's classical optimizer evaluates circuit candidates on an objective function and  picks the best circuits towards achieving the application's target.
Noise fluctuation can cause a significant transient impact on the objective function estimation of the VQA iterations' tuning candidates. This can severely affect VQA tuning and, by extension, its accuracy and convergence.

This paper proposes QISMET: \underline{Q}uantum \underline{I}teration \underline{S}kipping to \underline{M}itigate \underline{E}rror \underline{T}ransients, to navigate the dynamic noise landscape of VQAs. 
QISMET actively avoids instances of high fluctuating noise which are predicted to have a significant transient error impact on specific VQA iterations.
To achieve this, QISMET estimates transient error in VQA iterations and designs a controller to keep the VQA tuning faithful to the transient-free scenario.
By doing so, QISMET efficiently mitigates a large portion of the transient noise impact on VQAs and is able to improve the fidelity by 1.3x-3x over a traditional VQA baseline, with 1.6-2.4x improvement over alternative approaches, across different applications and machines.
%Further, to diligently analyze the effects of transients, this work also builds transient noise models for target VQA applications from observing real machine transients. These are then integrated with the Qiskit simulator.

\end{abstract}

%%
%% The code below is generated by the tool at http://dl.acm.org/ccs.cfm.
%% Please copy and paste the code instead of the example below.
%%


%%
%% Keywords. The author(s) should pick words that accurately describe
%% the work being presented. Separate the keywords with commas.
%% A "teaser" image appears between the author and affiliation
%% information and the body of the document, and typically spans the
%% page.

%%
%% This command processes the author and affiliation and title
%% information and builds the first part of the formatted document.
\maketitle

%%%%%%%%%%%%%%%%%%%%%%%%%%%%%
%sections from intro

\section{Introduction}

%%%%%%%%%%%%%%%%%%%%%%%%%%%%%%%%%%%%%%

%%%%%%%%%%%%%%%%%%%%%%%%%%%%%%%%%%%%%%%%%%%

%1. Why quantum is important
Quantum computers leverage superposition, interference, and entanglement to give them significant computing advantage in chemistry~\cite{kandala2017hardware}, optimization~\cite{moll2018quantum}, machine learning~\cite{biamonte2017quantum} and other domains of critical interest.
%2. NISQ
In near-term quantum computing, called Noisy Intermediate-Scale Quantum (NISQ), we expect to work  with  machines which comprise 100-1000s of imperfect qubits ~\cite{preskill2018quantum}. 

\begin{figure}[t]
\centering
\includegraphics[width=0.95\columnwidth,trim={0.5cm 0cm 0cm 0cm}]{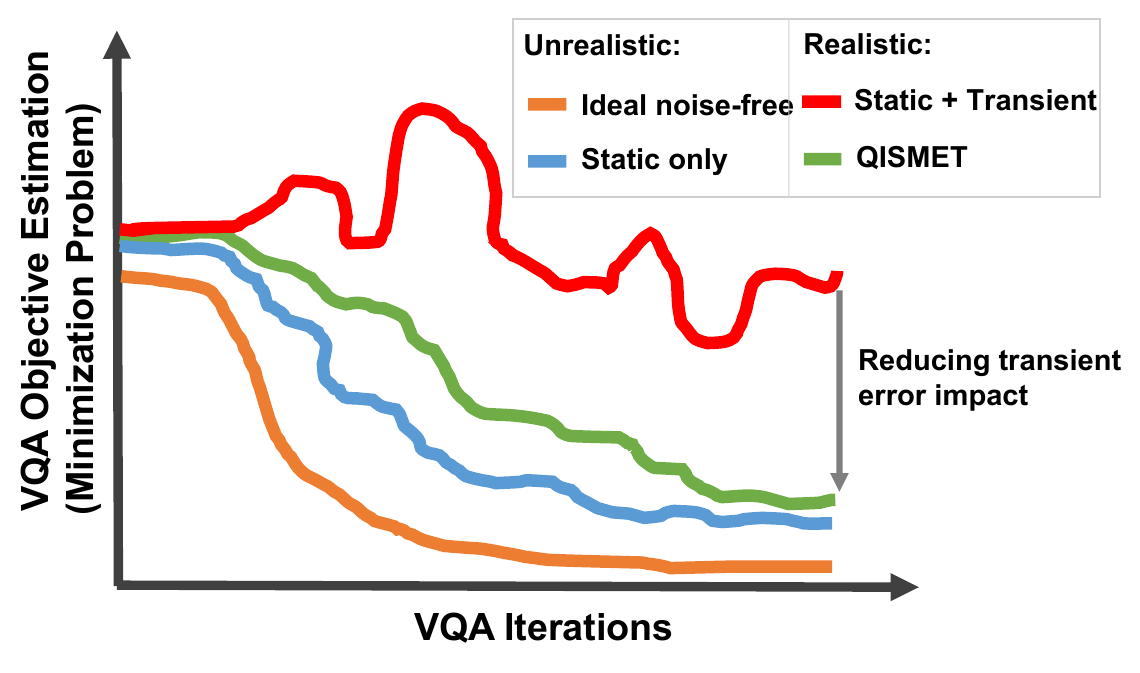}
\caption{%VQA benefits obtained with QISMET: 
Optimal VQA convergence is obtained in the ideal noise-free scenario (orange). In reality, VQA is affected by static and transient noise, and estimates can be much worse than ideal (red). QISMET (green) attempts to avoid significant transient error and thereby reaches close to the otherwise unrealistic blue line with only static noise. Prior error mitigation proposals can help bring the blue (and green) closer to the orange.
}
\label{fig:qismet_graph}
\end{figure}

%3. Transient errors in NISQ - what are they, static / transient, random / systematic, classical / quantum, external / internal, challenges in tackling them (needs a good background section
Today, noise prevents quantum computers from surpassing the capabilities of classical computers in almost all applications. 
NISQ devices suffer from high error rates in the form of state preparation and measurement (SPAM) errors, gate errors, qubit decoherence, crosstalk, etc.
These errors stem from multiple noise sources such as the imperfect classical control of the device, thermal fluctuations, destructive qubit coupling, imperfect insulation of the qubits, quasi-particles, and other external stimuli~\cite{muller2019towards,martinis2005decoherence,burnett2019decoherence,schlor2019correlating,Klimov_2018}.
The dynamic nature of quantum systems causes these noise sources to constantly vary over time, with some noise sources experiencing greater transient fluctuations than others. Additionally, because current fabrication techniques lack the precision to make homogeneous quantum device batches, the transient noise properties are unique to each device.

Effectively understanding and mitigating different forms of error is critical for quantum advantage in real-world applications.
Multiple error mitigation techniques have been explored on NISQ devices in the recent past~\cite{murali2019noise,tannu2019not,murali2020software,ding2020systematic, smith2021error,viola1999dynamical,pokharel2018demonstration, souza2012robust,temme2017error,li2017efficient,giurgica2020digital,tannu2019mitigating,bravyi2021mitigating}.
While these techniques have the potential to greatly improve execution fidelity, they almost always view machine noise as static, or at least expect the noise to be stable for sufficiently long periods of time such that their characteristics can be adequately captured, appropriately characterized, and then effectively mitigated.
Unfortunately, this is often insufficient---in this work we show that the transient errors from the dynamic noise landscape are challenging to comprehend and are especially detrimental to classes of applications that are iterative and/or long-running, and therefore their timely mitigation is important for significant fidelity improvements. 
{Here, we use the term `transient error' to refer to a temporary change in a quantum circuit’s output distribution, produced due to sudden non-insignificant shift in device characteristics of one or more qubits in the quantum circuit
~\cite{burnett2019decoherence,Klimov_2018,schlor2019correlating}. The effect of these transient errors is seen even with the statistical robustness offered by executing multiple circuit shots (which are required to capture probabilistic output distributions). }

The most popular examples of iterative long-running quantum applications are variational quantum algorithms (VQAs).
VQAs are considered one of the most promising quantum tasks for near-term quantum advantage in the NISQ era.
They have wide application in approximation~\cite{moll2018quantum}, chemistry~\cite{peruzzo2014variational} etc, which are usually designed as minimization problems.
VQAs are hybrid quantum-classical algorithms and they iteratively run a parameterized quantum circuit (QC) on the quantum machine.
The QC parameters are optimized each iteration by a classical tuner/optimizer to try and approach the global minimum of the variational objective function.
The classical optimizer inherently attempts to adjust the QC to the noise characteristics of the quantum device, and hence, \emph{in theory}, gives VQAs the potential for quantum advantage even on noisy machines.

Transient errors can be severely detrimental to VQAs.
Iteratively, the VQA tuner estimates gradients of some form~\cite{9259985} across multiple circuit candidates to choose the best set of VQA parameters for further tuning.
The VQA tuner works under the underlying assumption that the noise landscape of the device is unchanged during this gradient estimation process across the candidates.
This is often not the case. Noise fluctuation can cause a significant transient impact on the objective function estimation of one or more circuit candidates, and this can severely affect measured gradients and, therefore, the accuracy and convergence of VQA.

This discrepancy between VQA's ideal and reality is illustrated in Fig.\ref{fig:qismet_graph}.
The orange line shows the scenario with no noise, which is unrealistic (unless noise is entirely mitigated). 
The goal is to follow this line as closely as possible during VQA evaluation.
The blue line shows VQA estimations if only static noise was present (this could potentially encompass dynamic noise that varies at very coarse temporal granularity). 
This is again unrealistic, but most prior error mitigation and optimization proposals have focused primarily on this setting, attempting to lower this blue line to bring it closer to orange, achieving considerable benefits. 
The reality is represented by the red line which shows that VQA estimations are affected by both static as well as transient noise - traditional VQA optimizations can be less effective in dealing with transients.

This paper proposes \emph{QISMET: \underline{Q}uantum \underline{I}teration \underline{S}kipping to \underline{M}itigate \underline{E}rror \underline{T}ransients}, to navigate the dynamic noise landscape of VQA. QISMET tries to eliminate the effect of VQA transient errors to the maximum extent possible---lowering the red line in Fig.\ref{fig:qismet_graph} to bring it  close to the blue line (shown in green). 
Other error mitigation techniques can be added orthogonally to bring the blue line (and, by extension, QISMET's green line) closer to the orange ideal scenario.

QISMET is built with three key insights:
\circled{1} QISMET identifies a key capability in VQAs: a previous VQA iteration acts as an optimal reference circuit to estimate the effect of transient errors on a current VQA iteration. To utilize this, the current execution block reruns the reference circuit from the previous iteration, and the difference between the VQA outcomes is estimated.
\circled{2}\ Using traditional per-iteration gradient calculation, combined with QISMET's novel transient error estimation, QISMET is able to predict an estimate of the transient-free gradient for every iteration. 
\circled{3}\ Finally, QISMET designs a controller to decide if a particular iteration is acceptable to VQA or not. A VQA iteration is accepted only if the direction of the observed VQA gradient from the machine loosely matches the direction of the predicted transient-free gradient (relative to the previous iteration). In other words, a new set of VQA parameters is accepted only if the machine estimation of the parameters being good or bad, relative to the current parameters, matches the predicted transient-free estimation of the same (within some margin of error).

%4. VQA impacted NISQ and transient errors

%5. Proposal

%Contributions
\vspace{0.1in}
\textbf{\emph{Contributions and Results:}}

\circled{1}\ To the best of our knowledge, this work is among the first to study the effects of transient errors on VQAs.

\circled{2}\ This paper proposes QISMET, which actively avoids instances of high fluctuating noise which have a significant transient error impact on specific VQA iterations and can severely impact VQA accuracy and convergence.

\circled{3}\ To achieve this, QISMET estimates transient error in VQA iterations and designs a controller to keep the VQA iteration gradients faithful to the transient-free scenario.

\circled{4}\ By doing so, QISMET efficiently mitigates a large portion of the effects of transient noise on VQAs and is able to improve the fidelity of VQAs by 1.3x-3x over a traditional VQA baseline, across different applications and machines, {with 1.6-2.4x improvement over alternative approaches}.

\circled{5}\ To diligently analyze the effects of transients, this work also builds transient error noise models for target VQA applications from observing real device transients. These are then integrated with the Qiskit simulator.

\section{Background}

\noindent \emph{\textbf{Quantum Information:}}
\label{b_QI}
A qubit can exist in a linear superposition of the basis states $\ket{0}$ and $\ket{1}$, taking the form of $\ket{\psi}=\alpha\ket{0}+\beta\ket{1}$.
Together with other QC capabilities, this gives rise to the potential for quantum computers to exponentially outperform classical computers in certain applications.
Qubits are manipulated through gates that modify their probability amplitudes, i.e. the coefficients of $\ket{0}$ and $\ket{1}$. 
There are many non-trivial single-qubit gates that can modify qubit state. 
Pairs of qubits can be manipulated via multi-qubit interactions such as the two-qubit controlled-$X$, or $CX$ gate. 
Together, these enable universal quantum computation.

\emph{\textbf{Noise in the NISQ era:}}
\label{b_NISQ}
NISQ devices are error-prone and up to around 100 qubits in size today~\cite{preskill2018quantum}. %Environmental coupling is the source of many errors in quantum systems. 
These devices are extremely sensitive to external influences and require precise control, and as a result, some of the biggest challenges that limit scalability include limited coherence, gate errors, readout errors, and connectivity.
%The severity of each type of error varies per qubit and over time.
We will dive deeper into different noise sources and errors in Section \ref{motive}.
Multiple forms of error mitigation strategies have been proposed to correct different forms of quantum errors.
These include, but are not limited to, noise aware compilation~\cite{murali2019noise,tannu2019not};  correcting measurement errors~\cite{tannu2019mitigating,bravyi2021mitigating}; scheduling for crosstalk~\cite{murali2020software,ding2020systematic}; extrapolating for zero noise~\cite{giurgicatiron2020digital,li2017efficient,temme2017error, zne4}; decoherence mitigation through dynamical decoupling~\cite{DDBiercuk_2011,jurcevic2021demonstration,DDKhodjasteh_2007,DDPokharel_2018,souza2012robust}, spin-echo correction~\cite{hahn}, gate scheduling~\cite{smith2021error}; etc.
In addition, some of these can be used in conjunction to achieve better fidelity~\cite{ravi2021vaqem}.
Many of these techniques make deployment decisions that are dependent on noise characteristics of qubits and gates which have to be captured through expensive full machine characterization, and are thus captured only at some coarse granularity of time (eg., roughly once a day on IBMQ machines).
Therefore, they are suboptimal in the presence of fine-granularity dynamic fluctuations.

\begin{figure}[t]
\centering

\includegraphics[width=0.9\columnwidth,trim={0cm 0cm 0cm 0cm}]{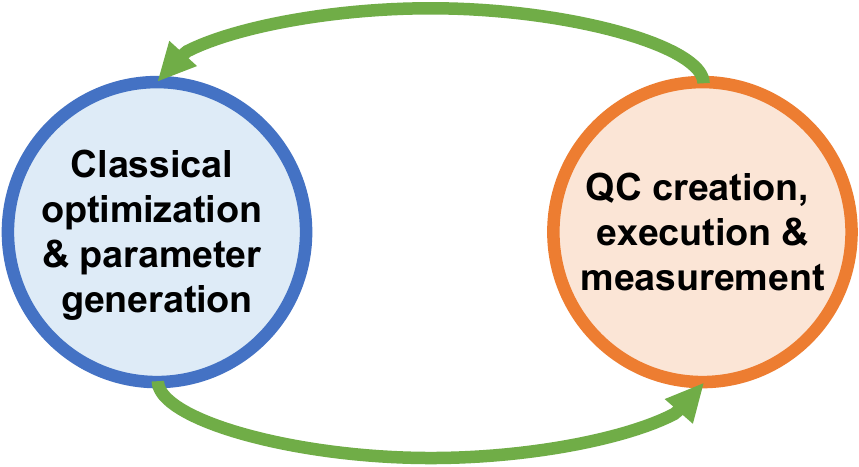}

\caption{VQA: a hybrid algorithm that alternates between classical optimization and quantum execution.}
\label{fig:vqa}
\end{figure}

\emph{\textbf{Variational Quantum Algorithms:}}
\label{b_VQA}
Variational algorithms expect to have innate error resilience due to hybrid alternation with a noise-robust classical optimizer~\cite{peruzzo2014variational, mcclean2016theory}. 
An overview of this process is illustrated in Fig.~\ref{fig:vqa}.
There are multiple applications in the VQA domain such as the Quantum Approximate Optimization Algorithm (QAOA)~\cite{farhi2014quantum} and the Variational Quantum Eigensolver (VQE)~\cite{peruzzo2014variational}.
Our applications in this work target VQE but QISMET is broadly applicable across all VQAs.
%VQE is used to find the lowest eigenvalue of a given problem matrix by computing a difficult cost function on the quantum machine and feeding this value into an optimization routine running on a CPU.
An important application of VQE is the ground state energy estimation of a molecule, a task that is exponentially difficult in general for a classical computer~\cite{Gokhale:2019}.
Estimating the molecular ground state has important applications in chemistry, such as determining reaction rates and molecular geometry.
At a high level, VQE can be conceptualized as a guess-check-repeat algorithm. 
The check stage involves the preparation of a quantum state corresponding to the guess.
%where the initial guess is often referred to as the ansatz. 
This preparation stage is done in polynomial time on a quantum computer, but would incur exponential cost  in general on a classical computer. 
This contrast gives rise to a potential quantum speedup for VQE~\cite{Gokhale:2019}.
The quantum circuit used in each iteration of VQE (and VQA in general) is termed an ansatz which describes the range of valid physical systems that can be explored and thus determines the optimization surface. 
Traditionally, the ansatz is parameterized by 1-qubit rotation gates. 
%A good ansatz provides a balance between a simple representation, efficient use of available native hardware gates, and sufficient sensitivity with the input parameters.
The VQA problem is represented as a Hamiltonian and is a linear combination of multiple Pauli terms.
The lowest eigenvalue of the Hamiltonian corresponds to the system's ground state energy~\cite{mcclean2016theory}.
Every iteration, the VQA objective function calculates the expectation value of this Hamiltonian.
This objective function is derived from ansatz measurements over different bases.
An illustration of this is shown in Fig.\ref{fig:qismet_design_qc}.

\emph{\textbf{Classical Tuning for VQAs:}}
\label{b_OPT}
The classical tuner/optimizer variationally updates the parameterized circuit until the measured objective converges to a minimum.
%This way it finds the corresponding eigenvalue and eigenstate. 
%Since VQE's Hamiltonian describes the energy evolution, this global minimum represents the ground state energy of the system. QAOA takes a similar approach.
For simple VQA problems and in the presence of minimal noise, the optimization surface is convex around the global minimum and smooth~\cite{9259985}. 
As noise increases, in line with NISQ machines, the optimization surface can potentially become non-convex and non-smooth. 
The surface contour worsens as the problem complexity increases because of increase in circuit depth, number of parameters, and entanglement spread~\cite{9259985}.
In this work we predominantly focus on the popular Simultaneous Perturbation Stochastic Approximation (SPSA)~\cite{SPSA} optimizer  which has shown some robustness to noise. 
SPSA is a method of stochastic gradient approximation, which only requires two measurements of the objective function per iteration, regardless of the dimension of the optimization problem.
%Limited evaluation with an alternative tuner called Implicit Filtering~\cite{9259985} is discussed in Section \ref{e_sim}.

\section{Motivation: Transient noise and its impact}
\label{motive}

\subsection{Device Level}
\label{m_device}

Non-uniformity among superconducting (SC) transmon qubits degrades the capacity of many near-term QCs to execute meaningful quantum workloads. 
At the physical level, every superconducting transmon qubit is unique because of the unavoidable process variation during device fabrication. 
These unintended differences across a quantum chip are called defects. 
One of the most severe defects that stochastically appears during transmon manufacture is a two-level system (TLS)~\cite{muller2019towards}. 
The name TLS might be confusing because a qubit, when working as intended, is also a system with two levels. However, the difference is that a TLS defect is inopportunely placed such that it destructively couples to qubits during computation, significantly reducing the duration of time in which quantum state information can be maintained~\cite{martinis2005decoherence}.

The Josephson Junction (JJ), or two superconductors separated by a thin metal-oxide insulator, is the critical component of the transmon qubit~\cite{gambetta2017building}. 
JJs are delicate components that must be fabricated with ultimate precision to achieve properties that enable the superconducting circuit to hold quantum state. 
Unfortunately, microscopic defects like TLS, taking the form of impurities inside materials or irregularities in atomic crystal structure, can appear in areas within the oxide layers or on the surface of the chip. 
TLS defects take on their own charge properties, and because each defect is unique, TLS is difficult to characterize. 
If TLS is close in proximity to the active components of the transmon, parasitic coupling causes qubit energy to be absorbed, causing fluctuation in qubit parameters such as T1 (amplitude) and T2 (phase) coherence times. 
Since transmon quantum systems are dynamic with evolving electric fields and states within JJs, the coupling strength of TLS that is near-resonant to a qubit varies over time, causing the qubit fluctuations to themselves be transient in nature~\cite{burnett2019decoherence,schlor2019correlating}. 
%Because of the significant influence TLS has on qubit coherence time, there have been many efforts to model and mitigate TLS influence on qubit T1 and T2.
Apart from TLS, qubit parameters are also affected (potentially to a lesser degree) by thermal fluctuations, magnetic flux, quasi-particles, etc.~\cite{Klimov_2018,burnett2019decoherence}.

\begin{figure}[t]
\centering
%\fbox{
\includegraphics[width=\columnwidth,trim={0.25cm 0.25cm 0.5cm 0.5cm},clip]{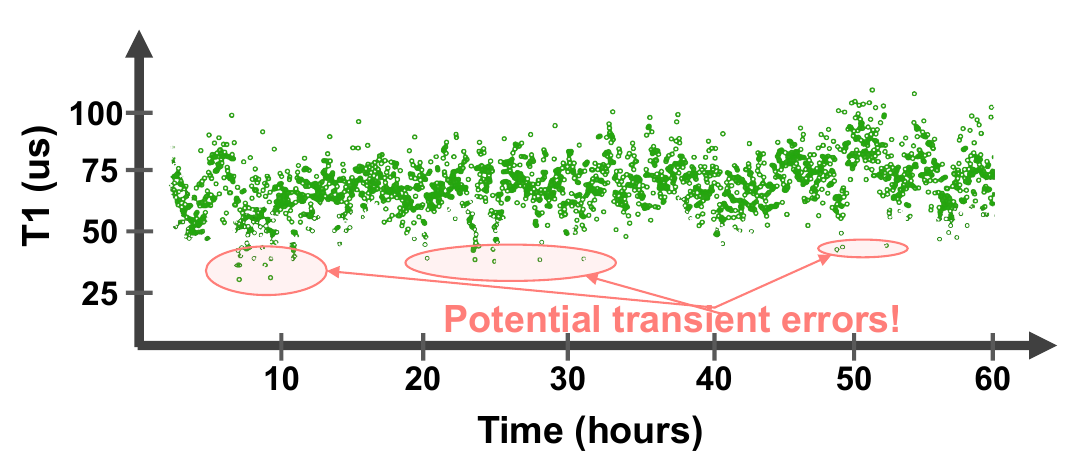}
%}
\caption{Transient fluctuations in T1 times observed over 65 hours~\cite{burnett2019decoherence}. %We expect circuits to suffer significant fidelity impact only from fluctuations resulting in very low T1 times (eg.,  circled instances).
}
%\caption{Transient fluctuations in T1 times observed over 65 hours on two transmon qubits~\cite{}. Fluctuations are primarily caused by unstable near-resonant two-level-systems. Not all fluctuations are detrimental to a target circuit. For instance, a low depth HW-efficient VQA ansatz might suffer significant detriment only for fluctuations resulting in T1 times less than 30us (left of dashed line). In this experiment, this occurs roughly 10\% and 1\% of the time for qubit Q1 and Q2 respectively (circled in orange) \note{update this fig: fred}. }
\label{fig:qismet_trans_qubits}
\end{figure}

Fig.\ref{fig:qismet_trans_qubits} shows transient fluctuations in T1 times observed over 65 hours on a transmon qubit~\cite{burnett2019decoherence} for reasons described above. 
It should be noted that not all fluctuations are detrimental to a target circuit.
We expect circuits to suffer significant fidelity impact only from fluctuations that result in very low T1 times.
These are the outliers that are circled in the figure.
Our expectation that impactful transients are an exception rather than the norm is evident in our analysis in the rest of this section and in our quantitative results.
Furthermore, other fluctuating qubit parameters can also have significant impact; T1 times are just one example.

%For example, a low-depth circuit such as a hardware-efficient VQA ansatz~\cite{kandala2017hardware} could suffer significant detriment only for noise fluctuations that result in T1 times less than 40us (below the dashed line).
%In this experiment, this occurs roughly 10\% of the time (circled).

\begin{figure}[t]
\centering
%\fbox{
\includegraphics[width=\columnwidth,trim={0cm 0cm 0cm 0cm},clip]{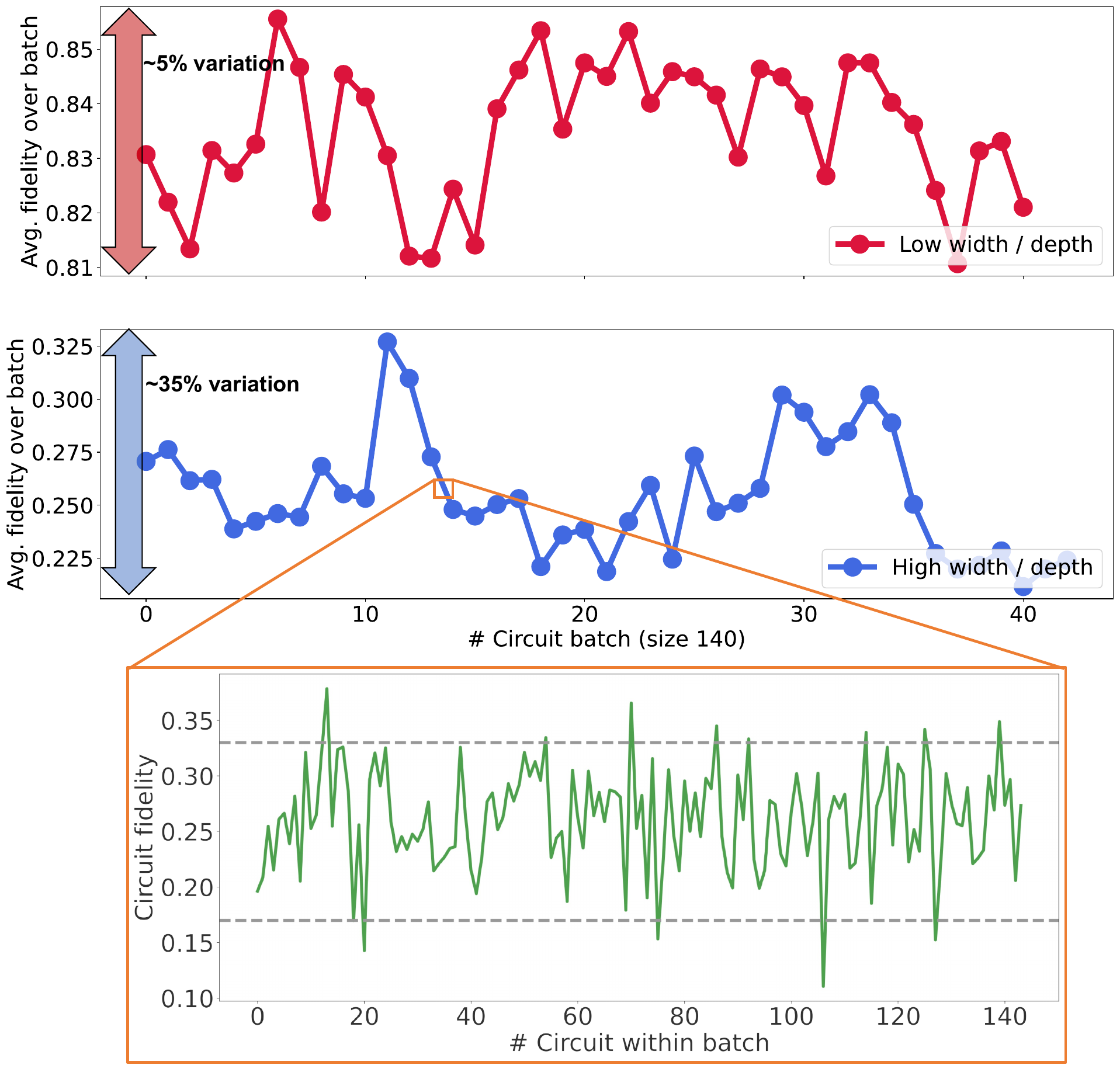}
%}
\caption{Impact of transient errors on circuits. The top two figures show circuit fidelity data collected over a 45-hour period. Each data point is the average circuit fidelity obtained over a batch of 140 (same) circuits over a one-hour period. The zoomed figure shows the variation in fidelity over one of the batches. 
%The transient error is observed to cause fine-grained fluctuations in circuit fidelity, and its impact is more significant in deeper and wider circuits. 
}
\label{fig:qismet_trans_circ}
\end{figure}

\subsection{Circuit Level}
\label{m_circuit}

Fluctuations in T1 times and other qubit/gate-level characteristics impact the execution of quantum circuits. The severity of impact could depend on: 

\noindent \circled{a}\ Width of the QC: More number of qubits increases the probability of high impact transients on some (or at least one) of the qubits, 

\noindent \circled{b}\ Depth of the QC: i) Deeper circuits are closer to the decoherence limit; therefore, a transient decrease in T1/T2 times can severely reduce circuit fidelity, and ii) deeper circuits usually imply more CX gates, which provide more sites for a substantial impact of error fluctuation,

\noindent \circled{c}\ Circuit State:  Effect of errors is state dependent. For example, a circuit that carries a superposition of states with a high proportion of 0s is less affected by decoherence.

Examples of the impact of transients on circuit fidelity are shown in Fig.\ref{fig:qismet_trans_circ}.
The two primary figures show circuit fidelity data collected over a 45-hour period. Each data point is the average circuit fidelity obtained over a batch of 140 circuits over a one-hour period.
The top circuit is a 4-qubit circuit with a depth of 6 CX gates. The average fidelity is around 83\% and the total variation in fidelity is only around 5\%.
The bottom circuit is 8-qubits with a depth of approximately 50 CX gates. The average fidelity is around 25\% and the variation is a concerning 35\%.
The zoomed figure at the bottom shows the fidelity variation over one of the batches, nearly 100\% variation across the batch. 
The dashed line indicates a 90\% threshold - in our applications we find that avoiding the extremes beyond this 90\% is the most beneficial trade-off.
More on this in Section \ref{e_thresh}.

\begin{figure}[t]
\centering
%\fbox{
\includegraphics[width=\columnwidth,trim={0cm 0cm 0cm 0cm},clip]{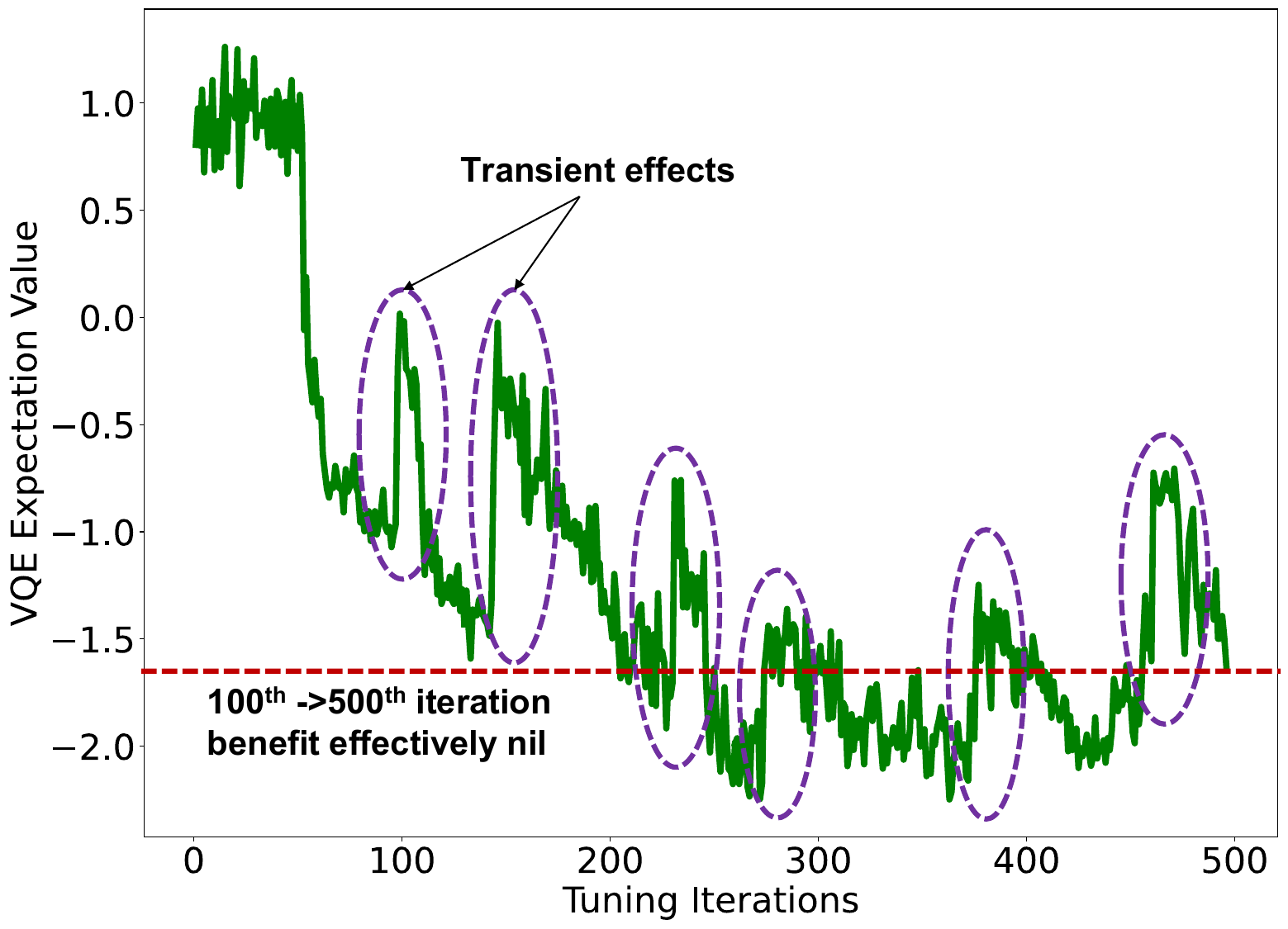}
%}
\caption{Extreme impact of transient errors on VQA tuning (multiple sharp spikes), from an experiment run on IBMQ Jakarta over a 24-hour period. %VQA convergence and accuracy are clearly impacted by transients.
}
\label{fig:qismet_trans_vqa}
\end{figure}

\subsection{Application Level (VQA)}
\label{m_app}

Finally, the impact of transient errors on quantum circuits have a cumulative effective on long-running applications such as VQAs whose outcomes are influenced by per-iteration circuit fidelity. 
Discrepancies in gradient evaluations caused by transient noise fluctuations can lead the classical tuner to proceed along unfavorable  directions in the VQA space, severely impeding convergence and accuracy.
Section \ref{p_navigate} discusses this in more detail.

%For every iteration of VQA, the classical optimizer evaluates gradients along different directions and then choose that which is most favorable towards attaining the global minima. 
%Achieving this is built on the assumption that all the gradients evaluated every iteration are affected by the same noise environment.
%Unfortunately, this is often not the case.
%As seen in previous sections, noise characteristics can vary widely even over fine granularities of time.
%Moreover, more complex VQA applications would in fact involve multiple circuit evaluations per iteration (for instance, for a Hamiltonian with a large number of Pauli terms) which would further spread out the evaluation of each gradient.

A severe case of transient error affecting VQA is shown in Fig.\ref{fig:qismet_trans_vqa}.
This is obtained from an experiment run on IBMQ Jakarta over roughly a 24-hour period. 
While we would ideally expect this VQA to somewhat monotonically converge to the minimum, multiple sharp spikes caused by transient fluctuations are seen. 
%While it is possible for the expectation values to increase when the optimizer attempts to jump out of a local minimum, the jumps are smoother and more moderate in height.
The spikes shown here are not the result of the optimizer jumping out of local minimas - such spikes are smoother and of lower amplitude.
Although the detrimental impact of some transient fluctuations might be negated over a few iterations (for instance, in the case of the first circled spike), others fluctuations have a more lasting impact (second circled spike).
The end expectation value reached at the end of 500 iterations is no better than that at 100 iterations, clearly showcasing the detrimental impact of transient noise. 
While such severe impacts might not always occur, milder scenarios are very common and are still detrimental, especially for long running use cases.

\begin{figure*}[t]
\centering
\includegraphics[width=0.95\textwidth,trim={-0.3cm 0cm -1cm 0cm},clip]{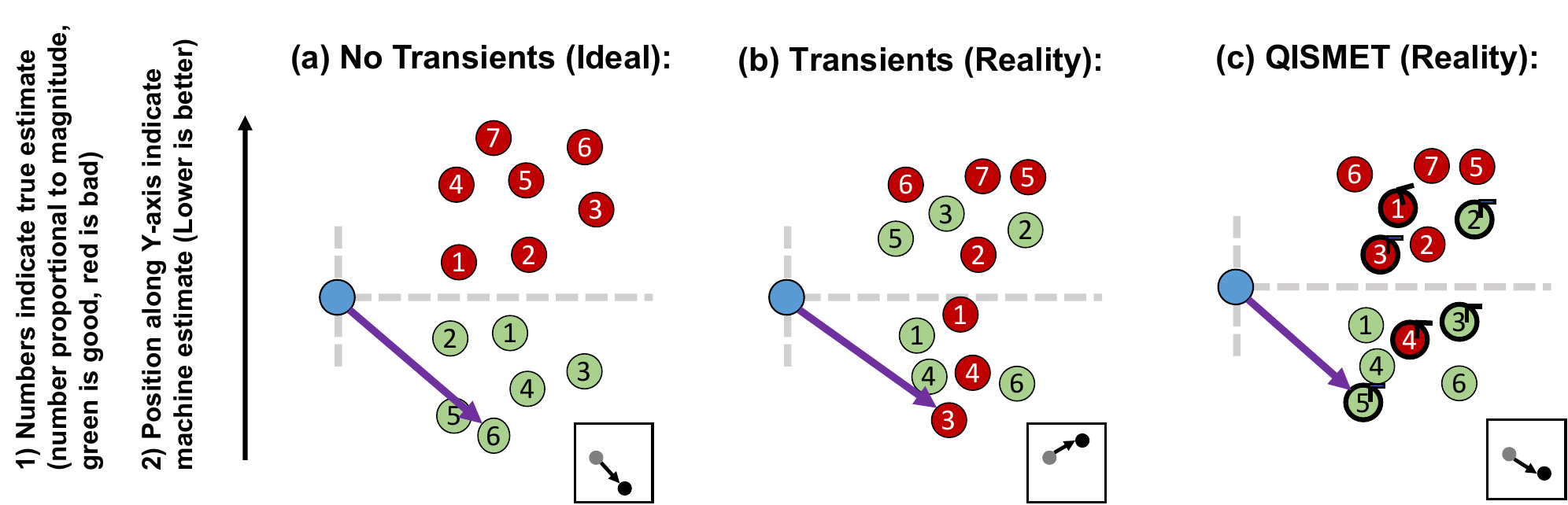}
\caption{QISMET navigating the VQA landscape. The pale blue dot represents the current configuration. Other dots represent different VQA configurations evaluated by the tuner. Y-axis positions indicate objective estimates (skewed by transients) as perceived by the VQA tuner - below the dashed line is perceived to be good. Numbers indicate true (transient-free) magnitude of objectives, while red and green indicate bad and good respectively. (a) In the ideal scenario, true and perceived estimates match and the tuner picks the true minimum (green-6). (b) In reality, the perceived minimum chosen by the tuner is a true bad configuration (red-3). (c) By estimating the transient effects, QISMET identifies some transient-affected configurations to be rerun. These are circled in black.  Rerunning these configurations enables most of them to escape significant transients, thus most true and perceived estimates get reasonably aligned. The perceived minimum is then chosen, which is green-5. The inset squares show effective VQA objective progress (moving lower is better).
}
\label{fig:qismet_intro_logic}
\end{figure*}

\section{QISMET Proposal}

\subsection{Navigating the Dynamic Noise Landscape}
\label{p_navigate}
In every iteration of VQA, a classical tuner evaluates the gradient along a particular direction.
Subsequently, it chooses directions which are some function of the prior evaluated gradients. 
Although different optimizers might implement this in various ways, most gradient-based approaches are some interpretation of the above.
For instance, in a simple steepest-path approach, the direction with the steepest gradient (towards the optimum) is chosen.
This tuning approach is built on the fundamental assumption that the direction gradients evaluated  are affected by the same noise environment.

Fig.\ref{fig:qismet_intro_logic} shows the VQA landscape for a minimization task under different scenarios.
The blue dot represents the current configuration. 
The other dots represent different VQA configurations evaluated by the tuner towards picking a subsequent configuration.
The dot positions along the Y-axis show their objective function estimates obtained on the machine in the presence of transients. 
Dots above the dashed line are \emph{perceived} by the VQA tuner to worsen the objective while those below are \emph{perceived} to improve it (`perceived' because the \emph{true} transient-free estimates can be different from the machine obtained estimates, which could be skewed by transients).
The color and number of each dot is indicative of its \emph{true} objective function estimate in the absence of transients - note that these true estimates are not simply known in reality.
Green dots improve the objective while red dots worsen it and the numbers indicate the magnitude.
In a steepest-gradient approach, the VQA tuner would ideally want to select the \emph{true} minimum - a green dot with highest magnitude number.
But in reality it picks the \emph{perceived} minimum  - i.e., the dot which is lowest in the vertical direction (indicated by arrow).
This setting is key to understanding the impact of transients on VQA and the QISMET solution.

\noindent \circled{a}\ The ideal scenario with no transients is illustrated in Fig.\ref{fig:qismet_intro_logic}.a.
Here, the configuration selected by the tuner, which is the lowest vertically, is, in fact, the green dot with highest magnitude (green-6).
This is because the true and perceived estimates are equal in an ideal transient-free setting.
Thus, it selects the true minimum among the evaluated samples.
The effective VQA objective progress is shown in the inset at the bottom (moving lower is better) - in this case, it is optimal.

\noindent \circled{b}\ Unfortunately, this is often not the case in reality as shown in Fig.\ref{fig:qismet_intro_logic}.b.
Transient noise induces discrepancies in the gradient evaluations, and this can cause the tuner to perceive configurations differently from their transient-free true estimates.
This is indicated in the figure by the colors/magnitudes not being correlated with vertical positions.
Some true good configurations are perceived as bad (green dots above the dashed line) and some true bad configurations are perceived as good (red dots below the dashed line). 
Furthermore, in this scenario, the perceived minimum which the tuner selects, i.e., the lowest dot in the vertical direction, is in fact a true bad configuration (red-3).
Choosing this configuration throws the VQA optimization significantly farther from the optimum, creating upward spikes as seen in Fig.\ref{fig:qismet_trans_vqa} and thereby derailing accurate convergence.
VQA progress is shown in the inset - the VQA objective worsens.

\noindent \circled{c}\ Fig.\ref{fig:qismet_intro_logic}.c shows how QISMET would navigate this realistic landscape with transient errors.
By estimating the transient noise (discussed in Section \ref{d_est}), QISMET tries to identify which perceived gradients are opposite in direction to their true gradients.
In other words, from Fig.\ref{fig:qismet_intro_logic}.b QISMET tries to predict which green dots are above the dashed line and which red dots are below.
It then re-executes these configurations on the quantum device until they are aligned with their true gradient directions (i.e., greens go below the dashed line and reds go above) or the retry budget is exhausted.
In Fig.\ref{fig:qismet_intro_logic}.c the retried configurations are circled in black.
Over the retrials, most (but not all) configurations realigned with their true gradient directions.
The realignment would happen for a configuration if it is able to execute in an instance of low transient noise - this is highly probable since transients of significant detrimental effect are fairly uncommon (as seen in Figures \ref{fig:qismet_trans_qubits} and \ref{fig:qismet_trans_circ}).
Eventually when selecting the new configuration, the tuner again selects the perceived minimum, i.e., the lowest vertical dot, but now this is a good true minimum (green-5), even if it is not the best (which was green-6 in the ideal case).
Since the selected configuration is still a good true minimum it does not derail the VQA optimization even in the presence transients.
VQA progress is shown in the inset - the VQA objective improves, even if not optimal (compared to the ideal case).
Thus, QISMET is more robust to the dynamic noise landscape of VQA.

\begin{figure*}[t]
\centering
\includegraphics[width=0.95\textwidth,trim={-3cm 0cm -3cm 0cm}]{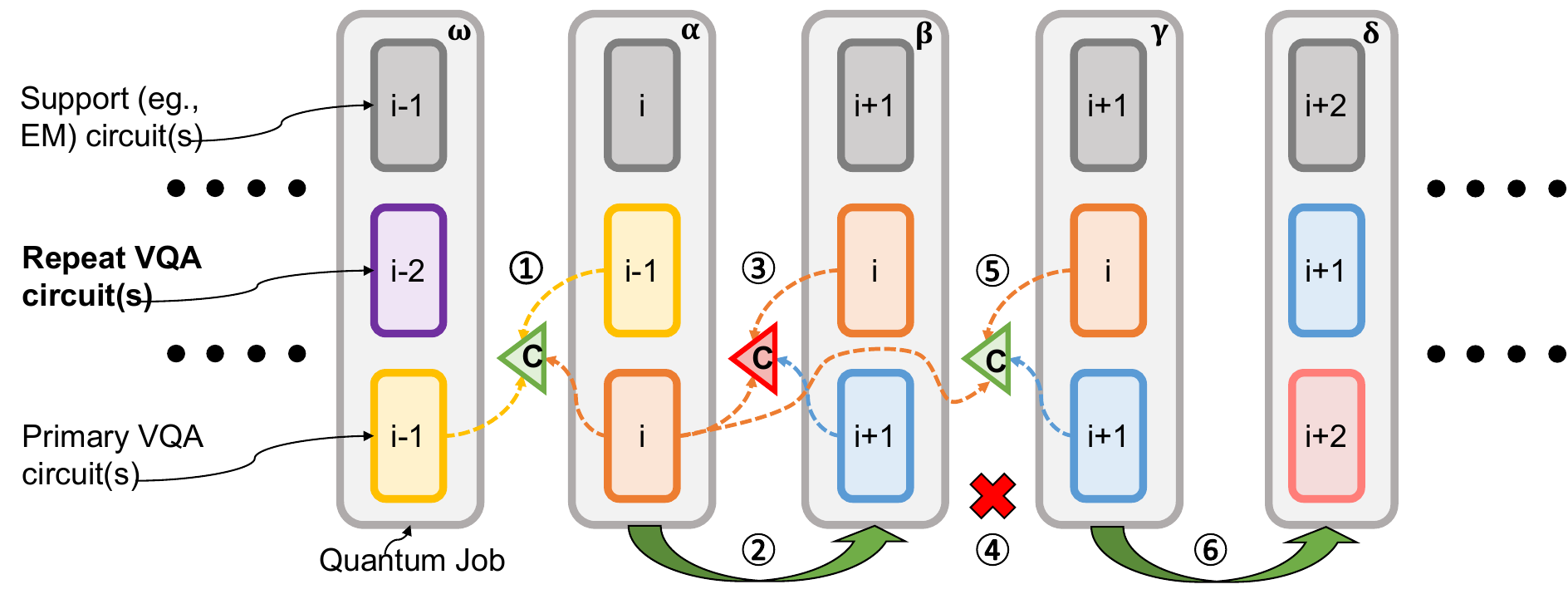}
\caption{Multiple VQA `jobs' (in light gray) run on the quantum machine. Job $\alpha$ consists of circuits corresponding to the $(i)^{th}$ VQA iteration (orange box), some error mitigation circuits (dark gray box), and QISMET adds circuits from the previous VQA iteration $(i-1)$ (yellow box). Other jobs/iterations can be inferred similarly from the figure. Progress in the VQA algorithm is determined by the QISMET controller (triangle labeled `C'). 
The following steps can occur: (1) After job $\alpha$, the QISMET controller takes as input the $(i-1)^{th}$ iteration circuits from previous job $\omega$ (yellow), the repeated $(i-1)^{th}$ iteration circuits in job $\alpha$ (yellow), and the $i^{th}$ VQA iteration from job $\alpha$ (orange). The controller then makes a decision that the transient noise impact on the $(i)^{th}$ iteration is reasonable. (2) VQA thus proceeds to the job $\beta$ with the primary circuit as the $(i+1)^{th}$ VQA iteration (blue), along with other circuits. (3) In this case, the controller deems that the transient noise impact is significant and can be detrimental to VQA. (4) Therefore, VQA does not proceed. Instead, the circuits are repeated via job $\gamma$. (5) When the controller checks again, it is deemed that the transient impact is within reasonable limits. (6) Thus, VQA proceeds to the next iteration and beyond.
}
\label{fig:qismet}
\end{figure*}

\subsection{Per-iteration QISMET Functionality}
\label{p_function}

Fig.\ref{fig:qismet} shows an instance of how QISMET functions over multiple iterations of VQA.
A sequence of VQA `jobs' are run on the quantum machine - each job is a collection of independent circuits.
%In the most minimal version of VQA, a job $\alpha$ (light grey outer box) that is run on the quantum machine would only consist of . 
If VQA is run with some error mitigation (measurement error mitigation, for example), job $\alpha$ would consist of circuits corresponding to the $(i)^{th}$ VQA iteration (orange inner box) and some related error mitigation circuits (dark gray inner box).
QISMET adds some reference circuits to job $\alpha$.
These are the repeated circuits from the  $(i-1)^{th}$ VQA iteration (yellow box).
Circuits in the other jobs shown can be similarly inferred from the figure.
Progress in the VQA algorithm is determined by the QISMET controller which is shown in the triangle labeled `C' between jobs. The following steps can occur:

\noindent \circled{1}\ Upon executing job $\alpha$, the QISMET controller takes as input: a) iteration $(i-1)$  circuits from the previous job $\omega$ (yellow), b) its repetitions in the current job, i.e., iteration $(i-1)$ circuits in job $\alpha$ (yellow), and c) the primary circuits of job $\alpha$, which correspond to iteration $(i)$ (orange). In this case, the controller is shown to make the decision that the impact of transient noise on the $(i)^{th}$ iteration is reasonable.

\noindent \circled{2}\ Thus, VQA proceeds to job $\beta$ with its primary circuit as VQA iteration $(i+1)$   (blue), along with other circuits, as shown in figure. 

\noindent \circled{3}\ After completion of job $\beta$, the QISMET controller now takes as input: a) the iteration $(i)$ circuits  from the previous job $\alpha$ (orange), b) its repetitions in the current job, i.e., iteration $(i)$ circuits in job $\beta$ (orange), and c) the primary circuit of job $\beta$, which correspond to iteration $(i+1)$ (blue). In this instance, the controller is shown to make the decision that the transient noise impact is significant and can be detrimental to VQA.

\noindent \circled{4}\ Therefore, VQA does not proceed. Instead the VQA iteration $(i+1)$ circuits along with  other circuits are repeated via job $\gamma$ (hence same coloring as job $\beta$). 
Note: since some transients effects can stay for extended time periods, the number of such rejections and repetitions is limited to some small max-out.

\noindent \circled{5}\ When the controller checks again (note that it obtains the iteration $(i)$  circuits in orange from job $\alpha$), it is deemed that significant fluctuation has passed and the current impact of transient noise is within reasonable limits.

\noindent \circled{6}\ Thus, VQA proceeds to job $\delta$ with the primary circuit as VQA iteration $(i+2)$, along with other circuits.

By taking steps as described above, the QISMET controller is able to navigate the dynamic noise VQA landscape, making decisions to the effect of Fig.\ref{fig:qismet_control_logic} as discussed in Section \ref{p_navigate}.
Transient error estimation is discussed in Section \ref{d_est} and controller logic in Section \ref{d_controller}.

\section{QISMET Design}

The QISMET proposal has two major components.
The first component estimates transient error in each VQA iteration and uses this to make transient-free predictions.
The second component uses the transient-free predictions to skip/retry particular iterations which are deemed harmful to VQA convergence.

\begin{figure*}[t]
\centering

\includegraphics[width=0.98\textwidth,trim={3cm 0cm 0cm 0cm},clip]{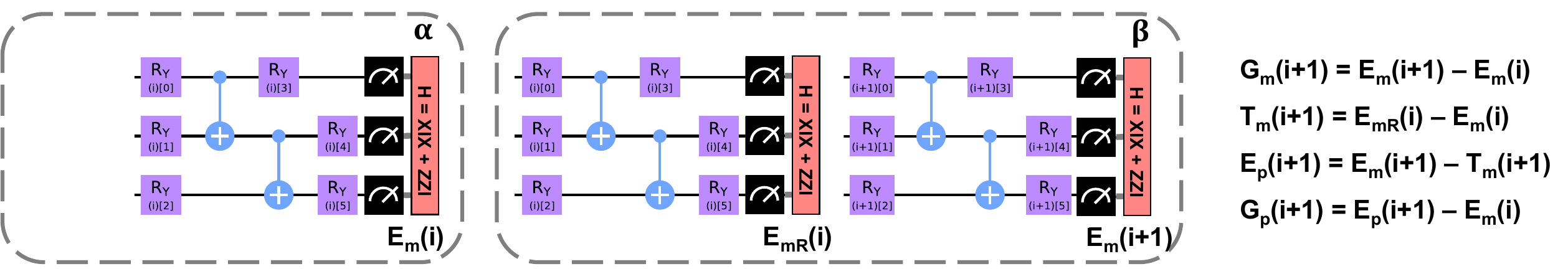}

\caption{Detailed expansion of Fig.\ref{fig:qismet}. VQA iteration $(i)$  in job $\alpha$ produces energy estimate $E_{m}(i)$. Iteration $(i+1)$ in job $\beta$ produces $E_{m}(i+1)$.  The rerun of iteration $(i)$ in job $\beta$ produces $E_{mR}(i)$. The gradient with transients, as observed by a traditional VQA tuner, is $G_{m}(i+1)$. QISMET estimates the machine transient noise on iteration ($i+1$) as $T_{m}(i+1)$ and then predicts transient-free  $E_{p}(i+1)$ and  $G_{p}(i+1)$. A VQA iteration is accepted by QISMET only if $G_{p}$ and $G_{m}$ point in the same direction as shown in Fig.\ref{fig:qismet_control_logic}. 
}
\label{fig:qismet_design_qc}
\end{figure*}

\subsection{Transient Estimation, Transient-Free Prediction}
\label{d_est}
Fig.\ref{fig:qismet_design_qc} is a detailed expansion of any single iteration of Fig.\ref{fig:qismet}.
Note that no error mitigation circuits from Fig.\ref{fig:qismet} are shown.
The left of Fig.\ref{fig:qismet_design_qc} shows the VQA iteration $(i)$ circuit that was run in the `previous' job $\alpha$. 
Its objective function (or energy) estimation on the machine is denoted by $E_{m}(i)$.
To its right is `current' job $\beta$ in which two circuits are shown.
The left circuit is the rerun of iteration $(i)$ and its energy estimation on the machine is denoted by $E_{mR}(i)$.
The right circuit is the iteration $(i+1)$ VQA circuit and its energy estimation on the machine is denoted by $E_{m}(i+1)$.
The machine gradient for iteration $(i+1)$ is $G_{m}(i+1)$ and is the difference between the machine energy estimates of iterations $(i+1)$ and $(i)$.
This is the gradient estimate that a traditional VQA tuner would use towards parameter selection for the next iteration.
QISMET estimates the transient error in iteration $(i+1)$ i.e., $T_{m}(i+1)$, as the energy difference between the rerun of iteration $(i)$ in job $\beta$ and its original run in job $\alpha$.
QISMET then predicts the transient-free energy for iteration $(i+1)$ i.e., $E_{p}(i+1)$, relative to $(i)$ by removing the transient error component from $E_{m}(i+1)$.
The predicted transient-free gradient is then $G_{p}(i+1)$.
Equations are shown to the right of Fig.\ref{fig:qismet_design_qc}.
This data is passed onto the QISMET controller which is discussed next in Section \ref{d_controller}.

Note that this assumes that the impact of the transient error on iteration $(i+1)$ is the same as that on iteration $(i)$. 
Although this is not necessarily exactly true, %the circuits between adjacent VQA iteration are usually close enough in their output states produced that this to be a fair assumption.
%In an alternate view, 
circuit $(i)$ is the closest possible reference circuit for transient noise estimation on circuit $(i+1)$, so its choice is justified.

\begin{figure}[t]
\centering
%\fbox{
\includegraphics[width=\columnwidth,trim={0cm 0cm 0cm 0cm},clip]{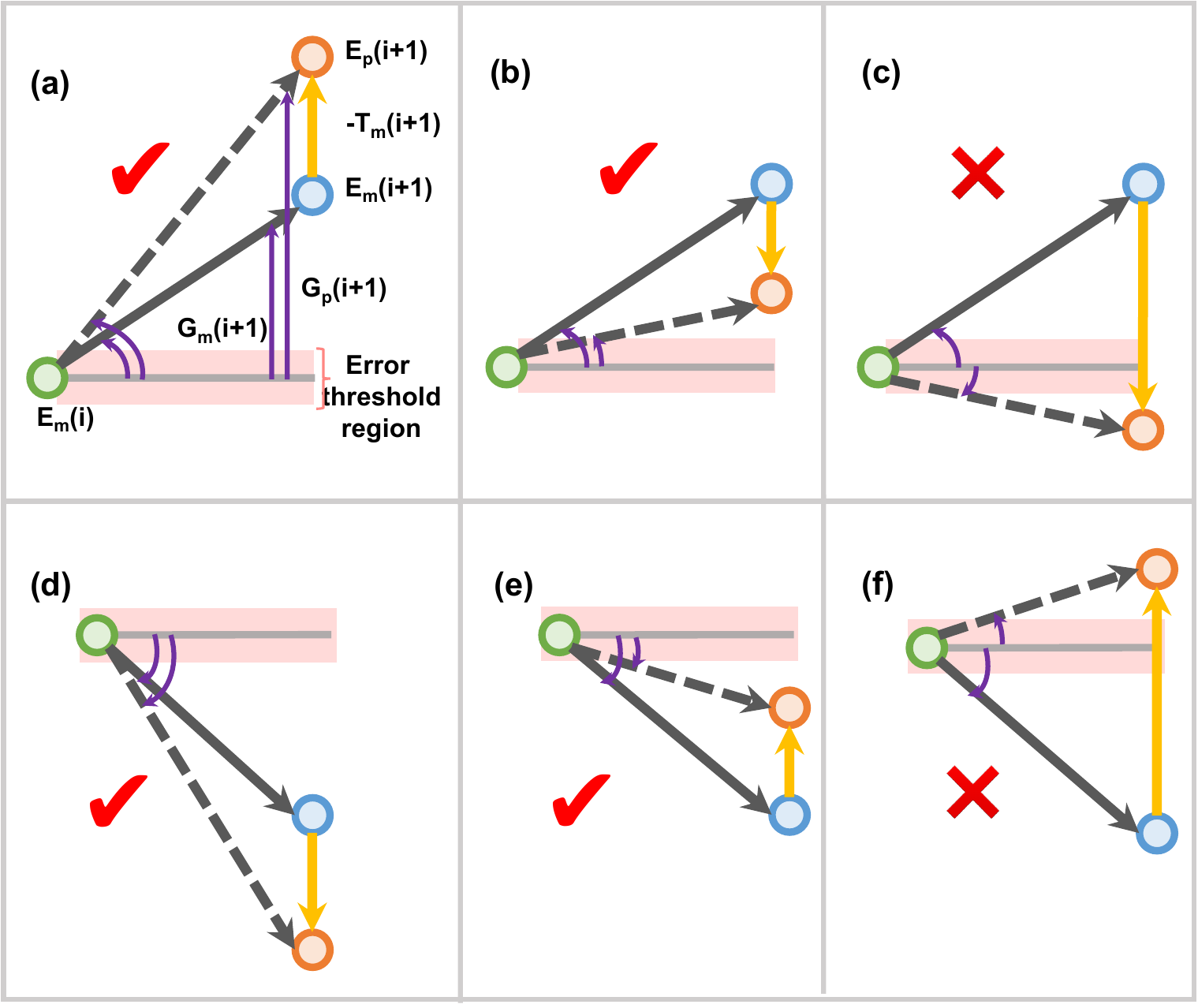}
%}
\caption{QISMET's gradient faithful controller. Green dots: VQA machine estimates for iteration $(i)$. Blue dots: machine estimate for iteration $(i+1)$. Yellow lines: negation of transient noise between $(i)$ and ($i+1$). Orange dots: predictions of transient-free estimates for iteration $(i+1)$. 
Difference scenarios and the reaction of the controller are shown.
(a) and (b): Both the machine and the transient-free estimates have a positive gradient. Thus, the estimates are acceptable. (d) and (e): Both machine and predicted transient-free estimates have a negative gradient, and are thus acceptable. (c) Machine estimate has a positive gradient but the predicted estimate has a negative gradient. This instance is unacceptable, since it makes a potentially bad VQA configuration be interpreted as a good one. (f) This is the opposite of (c) and is again unacceptable. Note that gradient swings within the shaded threshold region are always accepted.
}
\label{fig:qismet_control_logic}
\end{figure}

\subsection{Gradient Faithful QISMET Controller}
\label{d_controller}
Fig.\ref{fig:qismet_control_logic} is an illustration of different scenarios presented to the QISMET controller and the corresponding decisions made so that it achieves the navigation illustrated in Fig.\ref{fig:qismet_intro_logic}.

The intuitive idea is that the controller accepts VQA iterations only if the directions of the true gradients (i.e., gradients obtained in an ideal transient-free scenario) match the direction of the perceived gradients (i.e., gradients observed by the VQA tuner based on machine energy estimations).
This prevents scenarios such as truly bad VQA configurations being perceived as good configurations, which could throw the VQA optimization away from its path towards the optimum target.

In the figure, the green dots are VQA energy estimates made on the quantum machine for iteration $(i)$, i.e., $E_{m}(i)$.
The blue dots are machine energy estimates for iteration $(i+1)$, i.e., $E_{m}(i+1)$. 
The yellow lines are the negation of the estimated transient error between $(i)$ and $(i+1)$, i.e., $-T_{m}(i+1)$. 
They are used to produce the orange dots which are the predictions of the transient-free energy for iteration $(i+1)$ relative to $(i)$, i.e., $E_{p}(i+1)$.
Different scenarios and controller decisions are discussed below.

\noindent \circled{1}\ In Fig.\ref{fig:qismet_control_logic}.a and b, both the machine estimates and the predicted transient-free estimates for iteration $(i+1)$ have a positive gradient wrt $(i)$, i.e., $G_{m}(i+1)$ and $G_{p}(i+1)$ are both positive. Thus, the VQA iteration is acceptable, as the direction of the gradient is maintained.

\noindent \circled{2}\ In (d) and (e), the machine gradients and predicted transient-free gradients of $(i+1)$ wrt $(i)$ are both negative, thus this iteration is acceptable. 

\noindent \circled{3}\ In (c), the machine estimate for $(i+1)$  has a positive gradient wrt $(i)$, but since the transient noise is significantly large, the predicted estimate (after negation of the transient noise) has a negative gradient. This is unacceptable since it makes a potentially bad VQA configuration be interpreted as a good one. 

\noindent \circled{4}\ (f) is the inverse of (c) - again unacceptable because it makes a good VQA configuration be interpreted as bad. 

\noindent \circled{5}\ Finally, the gradient swings within a specified error threshold region (pink shaded region) are always accepted. This is to avoid frequent skipping on less impacting transients. This is discussed and evaluated in Section \ref{e_thresh}. 

\subsection{Other Skipping and Filtering Techniques}
\label{filter}
It could be argued that an alternative intuitive solution is to avoid all instances in which transient noise is greater than some set threshold. 
The controller could simply be designed to skip a VQA iteration $(i)$ if abs($T_{m}(i)) > \tau$.
We find that such a solution is unsuitable because many transients (even of reasonable magnitude) that are constructive to VQA progress, are skipped, thus, affecting accuracy and substantially delaying convergence.
A quantitative comparison is discussed in Section \ref{e_alt}.

{
The above is a fundamental difference between the QISMET gradient faithful controller and other filtering techniques that are common in classical computing (such as in signal processing).
For example, Kalman filtering~\cite{kalman} is a signal processing approach that uses a system's environment model, some known inputs to that system, and multiple noisy measurements (say, from sensors), to build an understanding of the system's noise/variation, and then calculate noise-free estimates.
It is generally suitable to application settings with dynamic variations similar to our focus in this work.
A key feature of such filtering techniques is that they treat all forms of measurement variance to be the same and eliminate the variance in a principled way by inferring some information from multiple measurements.}

{
Unfortunately, this is less suited to the VQA tuning landscape.
As discussed in Section \ref{d_controller}, not all transients are harmful to VQA progress and only those that flip the gradient directions are detrimental and need to be avoided. 
From Fig.\ref{fig:qismet_control_logic}, let's consider scenarios (a) and (c).
While both have significant transient error magnitudes, (a) is not detrimental to VQA progress and is accepted by QISMET while (c) is deemed detrimental and is rejected.  
In contrast, the Kalman Filter (and related techniques) will
incorporate some information from both (a) and (c) towards its elimination of noise and estimation of the true value.
When carefully tuned, the differences  in these scenarios can be incorporated into the filter to some extent, but its efficiency will be limited.
With less careful tuning, the filtering might, in fact, prove to be detrimental. 
We quantitatively showcase different filtering scenarios in Sec.\ref{kalman}.}

{
It should also be noted that Kalman and other classical filtering techniques work best when the signal and noise follow some specific fairly stable characteristics.
%data fits well into specific fairly stable distributions.
%the noisy nature of the data can be well represented mathematically and this representation is relatively stable over the runtime of the application.
%This is possible in the classical world, where sensor data can be obtained inexpensively in many classes of applications and environment noise under ambient conditions is relatively stable.
%Unfortunately, quantum data is expensive to collect, both in terms of time overhead and in monetary cost~\cite{gu2021adaptive}.
%Expensive estimates can sometimes be replaced by surrogate approximation~\cite{zamora2021} but designing a faithful quantum surrogate is a separate research endeavor in itself. %TODO CAMERA
Unfortunately, as per our current understanding of quantum devices, the transient noise dynamism is largely non-deterministic and constantly varying over fine granularities of time.
}

%In the absence of sufficient data, coupled with highly non-deterministic device noise, the benefit from classical filters such as Kalman is limited. 

\section{Methodology}
\label{6-method}

\begin{table}[]
\caption{TFIM VQA applications for simulation}
\resizebox{\columnwidth}{!}{
\begin{tabular}{|l|l|l|l|l|}
\hline
\textbf{Application} & \textbf{Qubits} & \textbf{Ansatz} & \textbf{Reps} & \textbf{Machine + trial} \\ \hline
\textbf{App1} & 6 & SU2 & 2 & Toronto (v1)   \\ \hline
\textbf{App2} & 6 & RA  & 4 & Guadalupe (v1)  \\ \hline
\textbf{App3} & 6 & RA  & 4 & Guadalupe (v2)  \\ \hline
\textbf{App4} & 6 & SU2 & 4 & Toronto (v2)   \\ \hline
\textbf{App5} & 6 & RA  & 8 & Cairo (v1)     \\ \hline
\textbf{App6} & 6 & RA  & 8 & Casablanca (v1) \\ \hline
\end{tabular}
}
\label{Table1}
\end{table}

\subsection{Applications and General Infrastructure}

QISMET is a software optimization that can be integrated with any VQA classical optimizer.
We implement it in Python and use it with the Qiskit VQA framework~\cite{Qiskit}, through which it interacts with quantum execution.

We limit ourselves to one VQA domain, the VQE, which was introduced in Section \ref{b_VQA}.
Due to noisy machine limitations on circuit width and depth, applications are restricted to 6 qubits. 
Our evaluations encompass 2 Hamiltonians, 6 different ansatz, and 7 IBMQ machines (as well as simulation).
We primarily use the SPSA classical tuner across all our evaluations.
%, except for the discussion at the end of Section \ref{e_sim} which uses ImFil.
%These were discussed earlier in Section \ref{b_OPT}.

Our primary focus is the one-dimensional TFIM (Transverse Field Ising Model) Hamiltonian, which is an ubiquitous model that has applications in understanding phase transitions in magnetic materials~\cite{uvarov2020machine}. 
The TFIM is a desirable system since it is exactly solvable via classical means.
Further, the Hamiltonian is easily constructed to create small problem sizes.
We also evaluate the potential energy of the Hydrogen molecule over bond lengths of 0.4 - 2 \AA. 

We use the hardware efficient SU2~\cite{IBM-SU2} and RA~\cite{IBM-RA} ansatz which are of low depth and therefore suited to NISQ devices. 
%The ansatz is constructed for `circular' and `full' entanglements, which are different techniques to add $CX$ gates to the circuit.
The number of block repetitions in the ansatz is varied between 2, 4 and 8 repetitions.
IBM Q machines targeted for machine runs are Guadalupe (27q), Toronto (27q), Sydney (27q), Casablanca (7q), Jakarta (7q), and Mumbai (27q).
IBM Q machines from which traces are generated for simulation are Guadalupe (27q), Toronto (27q), Cairo (27q), and Casablanca (7q).
Machine details can be found on the IBM Quantum Systems page~\cite{IBMQS}.
TFIM simulation application details are summarized in Table \ref{Table1}.

\subsection{Building Transient Noise Models for Simulation}

While execution on real quantum machines is a must for the holistic evaluation of application fidelity, sufficient access to machines is still limited today, especially for long-running applications such as VQAs.
Further, fair evaluation requires a deterministic environment across different points of comparison, as well as reproducible results to validate and improve upon the proposal. 
To enable longer VQA runs in a deterministic environment and for reproducible comparisons, we are required to explore the use of IBM's quantum simulator (via Qiskit).
Unfortunately, the noise models traditionally available to Qiskit are static over the period of a machine's calibration cycle - they are refreshed roughly once per day.
%But, as was discussed in Section \ref{motive}, transient noise fluctuates at a machine finer granularity.
%Thus, existing simulator noise models are insufficient to study transient noise effects.

To enable fine-granularity transient noise analysis in simulation, we build transient error traces for our target application-machine combinations.
Per-iteration transient effects on VQA are captured and normalized to the magnitude of the VQA estimations.
These transient effects are composed into a data structure and integrated into Qiskit's VQA framework.
In each simulated VQA iteration, an instance of transient noise is accessed from the data structure corresponding to the particular application-machine pair.
This transient noise is factored in with existing static components from Qiskit's noise models.

\begin{figure}[t]
\centering
%\fbox{
\includegraphics[width=\columnwidth,trim={0cm 0cm 0cm 0cm},clip]{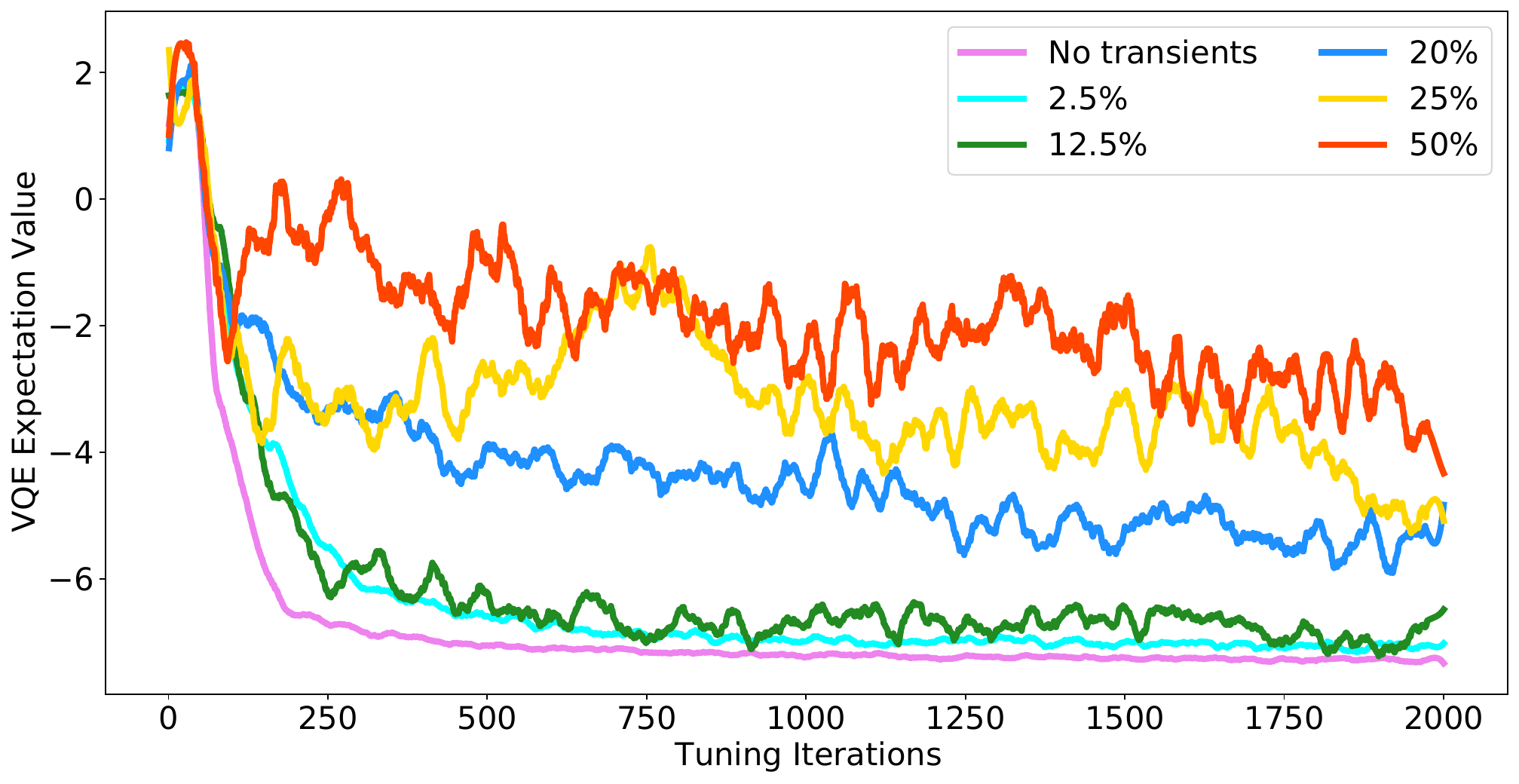}
%}
\caption{VQA simulation on the Qiskit simulator with our added transient noise model. 
%In our model, the transient noise and their impact on VQA can be varied, allowing for experiments like the above which increases the transient noise from 0 to 50\% of the ideal VQA objective estimations. As transient noise impact  increases, the accuracy and convergence of VQA estimations worsen.
}
\label{fig:qismet_sim_sweep}
\end{figure}

Furthermore, the magnitude of transient noise can also be varied to perform a wider range of analyzes, as shown in Fig.\ref{fig:qismet_sim_sweep}.
The figure shows a variety of transient noise magnitudes applied to a VQA application, varying from 0-50\% of the ideal VQA objective estimations. 
Intuitively, it is evident that as the impact of transient noise increases, the accuracy and convergence of the VQA estimates worsen.
Similar experiments can be performed to test the capability of new proposals - we use this simulator to extensively evaluate QISMET.

\subsection{Evaluation Comparisons}
\label{eval_cmp}
We evaluate QISMET based on VQA energy estimates (using the standard Hartree Energy metric when applicable). %Note that this is an application-level quality metric which is relevant to VQA, as opposed to circuit-level fidelity metrics that are useful for general quantum circuits.
In one or more experiments, we compare:

\emph{Baseline:} Traditional VQA employing measurement error mitigation but no mitigation for transient errors.

\emph{QISMET:} Baseline + control to estimate transients and skip/rerun iterations when appropriate. The QISMET error threshold is set so as to skip at most 10\% of the iterations (found to be the most suitable - Section \ref{e_thresh}).

\emph{Blocking:} A Qiskit SPSA optimization that only accepts VQA updates that move towards the objective. 

\emph{Resampling:}  SPSA optimization that increases the number of times the gradient is sampled (we use 2x). 

\emph{2nd-order:} A Qiskit SPSA optimization that estimates second-order derivatives to condition the gradient. %Under transient noise, the second order derivatives can be skewed as well, and can further worsen VQA.

\emph{Noise-free:} An ideal scenario without any form of error (run on the Qiskit simulator).

\emph{QISMET-conservative:} QISMET with threshold set to skip at most 1\% of the VQA iterations.

\emph{QISMET-aggressive:} QISMET with threshold set to skip at most 25\% of the VQA iterations.

\emph{Only-transients:} An alternative proposed technique that skips VQA iterations if the estimated transient error is larger than some specified threshold.

{\emph{Kalman:} A classical filtering approach discussed in Section \ref{filter}. It requires tuning filter hyper-parameters such as the Transition Coefficient (T), which is a linear estimation of the slope of the noise-free curve, and the Measurement Variance (MV), which is related to the variance in the noisy measurements. 
}

\section{Evaluation Results}

\begin{figure}[t]
\centering
%\fbox{
\includegraphics[width=\columnwidth,trim={0cm 0cm 0cm 0cm},clip]{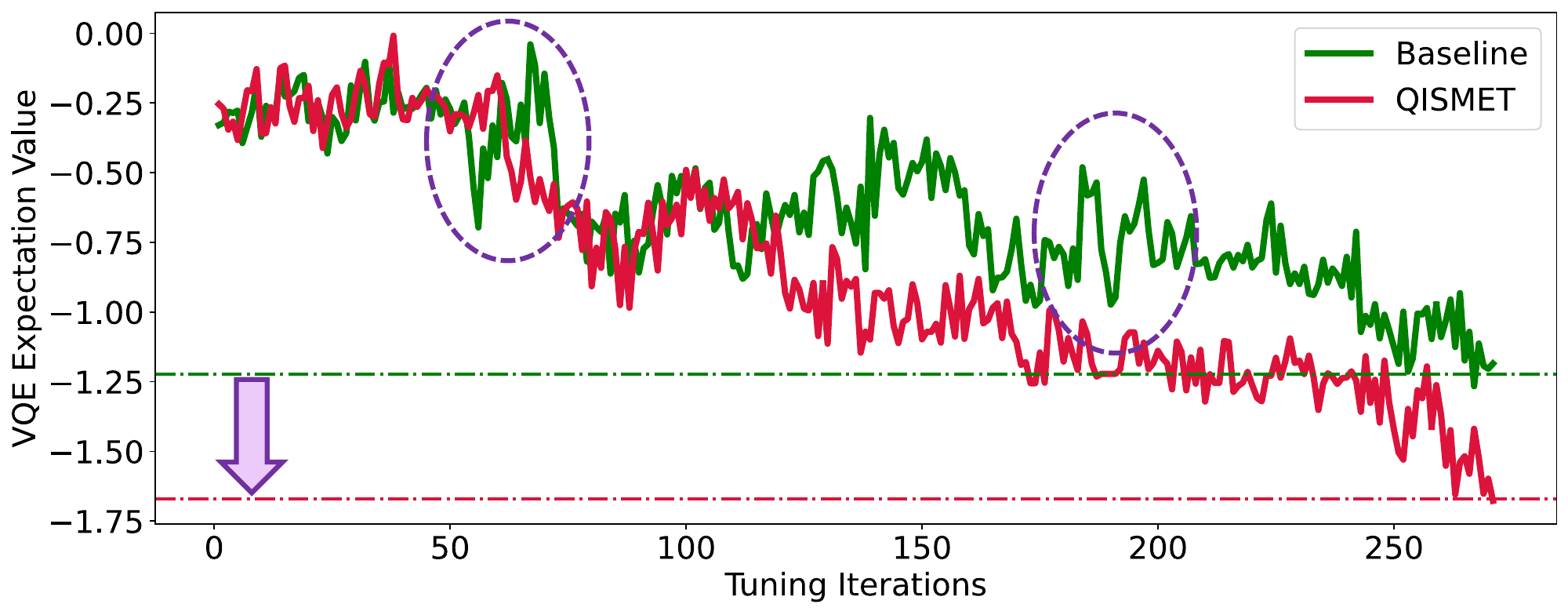}
%}
\caption{QISMET benefits for a 6-qubit TFIM VQA on IBMQ Guadalupe run over a 48-hour period. Two instances of moderate transient error are circled and are avoided by QISMET.
%These transients are avoided with QISMET, leading to a 40\% improvement in VQA estimation.
}
\label{fig:qismet_machine_guad}
\end{figure}

\begin{figure}[t]
\centering
%\fbox{
\includegraphics[width=\columnwidth,trim={0cm 0cm 0cm 0cm},clip]{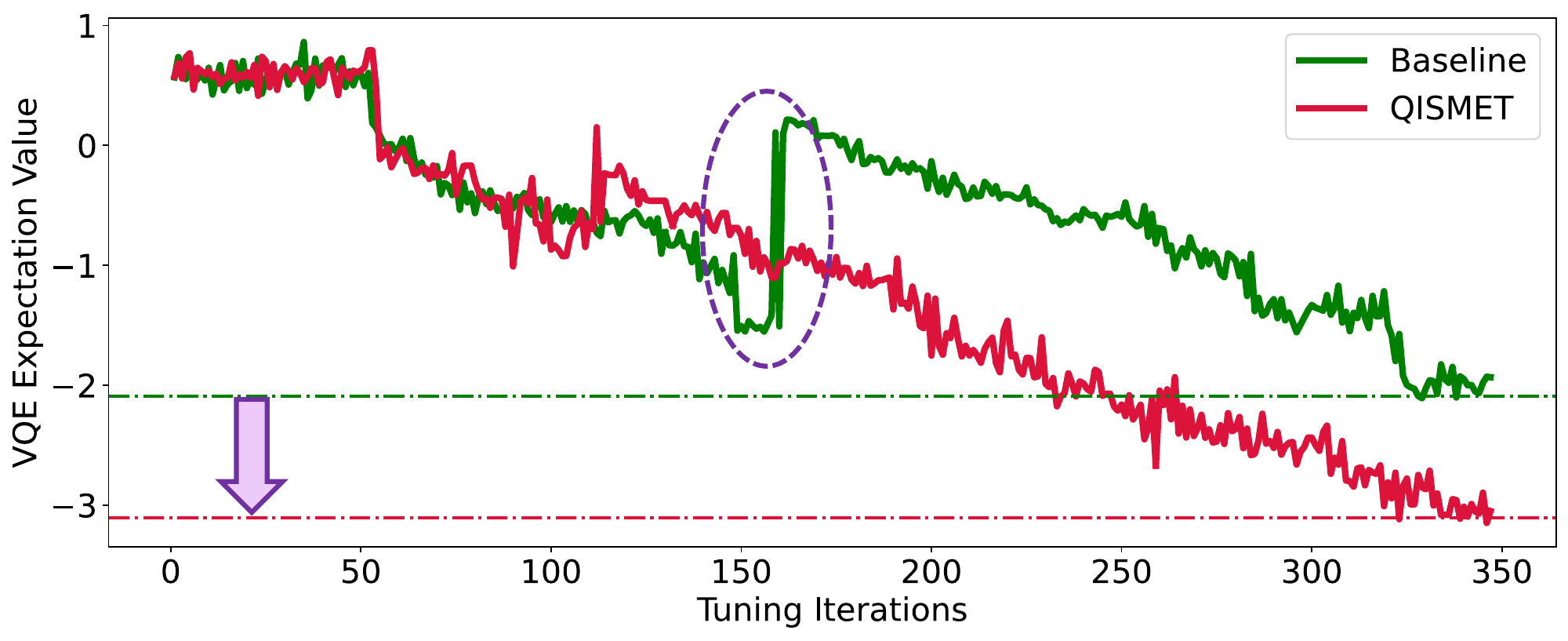}
%}
\caption{QISMET benefits for a 6-qubit TFIM VQA on IBMQ Sydney run over a 48-hour period with one sharp instance of transient error (circled) and are avoided by QISMET.
%is seen to considerably affect the baseline. This is avoided with QISMET, leading to a 50\% VQA improvement.
}
\label{fig:qismet_machine_syd}
\end{figure}

\begin{figure}[h]
\centering
%\fbox{
\includegraphics[width=\columnwidth,trim={0cm 0cm 0cm 0cm},clip]{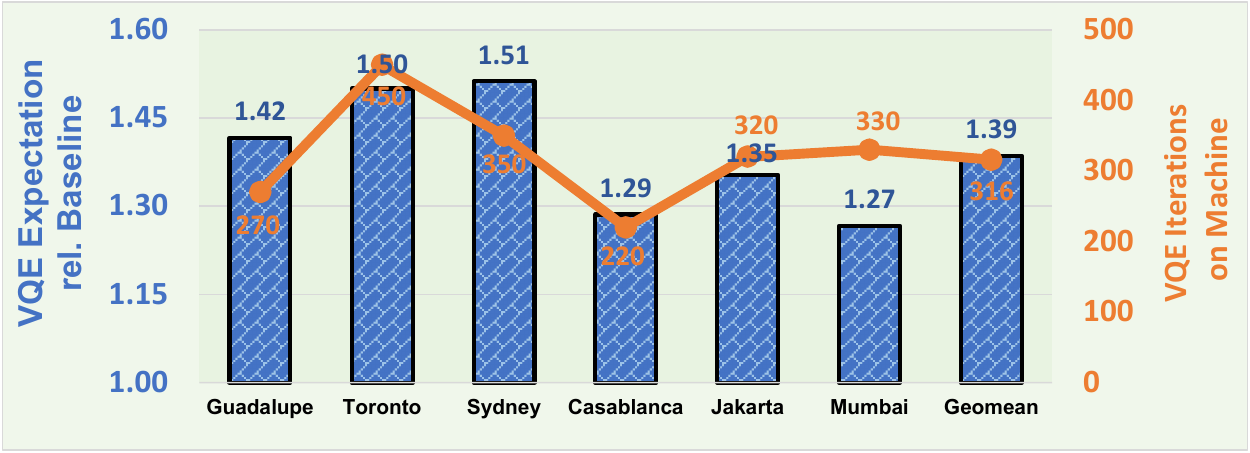}
%}
\caption{QISMET benefits for a 6-qubit TFIM VQA on six IBMQ machines. Iterations varies across applications, depending on machine availability. 
%On average, 40\% VQA improvement is observed with QISMET in around 300 iterations - more iterations will increase QISMET benefits.
}
\label{fig:qismet_machine_full}
\end{figure}

\subsection{Real Machine Experiments}
First, we analyze VQA experiments run on the real IBMQ quantum machines. Note that the number of iterations that we can run on these machines is limited due to access constraints.

Fig.\ref{fig:qismet_machine_guad} shows a comparison of QISMET against the baseline on a 6-qubit TFIM VQA on IBMQ Guadalupe. 
Both versions of the application are run for around 270 iterations, over a 48 hour period. 
The versions are run in a synchronous manner, so each iteration of the baseline is run temporally adjacent to the corresponding iteration of the QISMET version. 
Many instances of moderate transient error impact the baseline - two phases with transient errors are circled. 
While the baseline quickly recovers from the first transient phase, the second phase is more harmful, causing the VQA accuracy to somewhat stagnate over the next 50-100 iterations.
These transient phases are predominantly avoided with QISMET, leading to a 40\% improvement in VQA estimation over 270 iterations.

Fig.\ref{fig:qismet_machine_syd} shows a comparison of QISMET against the baseline for the same 6-qubit TFIM VQA  on IBMQ Sydney.
Both versions of the application are run for around 350 iterations, over a 48 hour period. 
%The versions are run in a synchronous manner, so each iteration of the baseline is run temporally adjacent to the corresponding iteration of the QISMET version. 
While the machine behavior is smooth for most of the tuning period, there is a single phase of significant  turbulence (circled) which heavily impacts the baseline.
%After a sharp back and forth, the baseline settles in a different region of the VQA space and continues its pursuit towards minimum, but with nearly 150 iterations of no effective improvement. %TODO CAMERA
QISMET is able to avoid this turbulent phase - QISMET is especially effective here since the transients only occur for a few iterations - once QISMET skips those errors, it is able to continue its steady progress, achieving a 50\% improvement over the 350 iterations.

Fig.\ref{fig:qismet_machine_full} shows the QISMET benefits over the baseline on the primary vertical axis and the number of iterations run on the secondary vertical axis.
Results are shown across 6 different IBM Q machines.
QISMET improves the measured VQA expectation by by 29-51\% across the machines, with a mean improvement of 39\%.
These improvements are achieved over 200-450 VQA iterations run on the devices.
We expect that the relative QISMET benefits would be greater if more iterations were run since there would be increased potential for more transients to impact the baseline.
%This is reasonable to expect and in line with the observation in Fig.\ref{fig:qismet_machine_syd} that even a single tumultuous execution phase can considerably throw VQA away from its optimal path of optimization.

\begin{figure}[t]
\centering
%\fbox{
\includegraphics[width=\columnwidth,trim={0cm 0cm 0cm 0cm},clip]{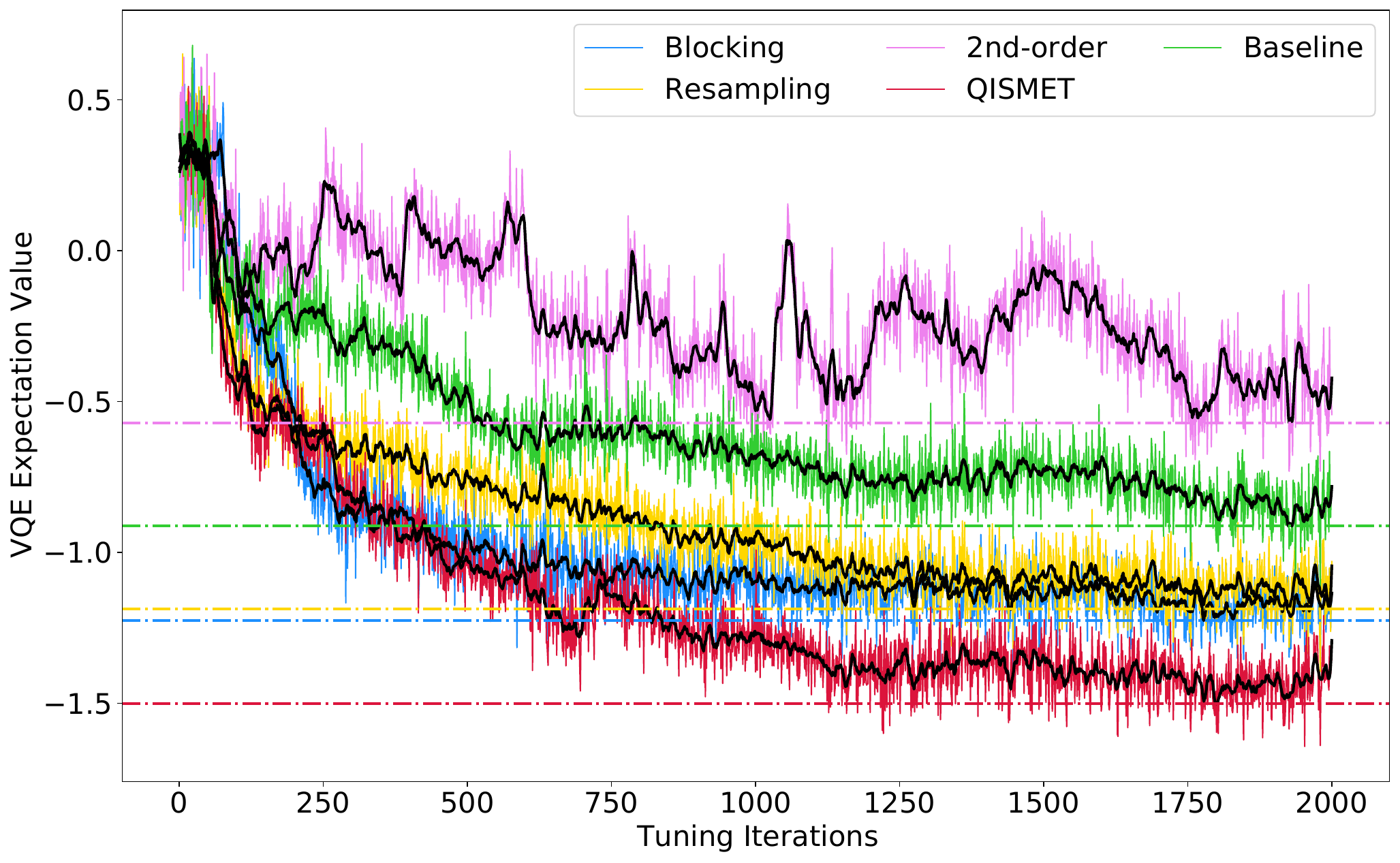}
%}
\caption{Simulating VQA with transient errors for App2, using the SPSA tuner, over 2000 iterations. QISMET is compared against SPSA optimizations.
%The Blocking and Resampling schemes offer some robustness to transient errors, but are very slow to converge to high accuracy estimates. The 2nd-order  scheme is considerably worse than the baseline.
%QISMET offers a 65\% improvement over the baseline.
}
\label{fig:qismet_sim_app}
\end{figure}

\subsection{Comparing against SPSA Optimization Schemes}
\label{spsa_cmp}

Fig.\ref{fig:qismet_sim_app} shows simulated evaluations of App2 (see Table \ref{Table1}) for five schemes - Baseline, QISMET, Resampling, Blocking and 2nd-order.
These are introduced in Section \ref{6-method}.
Simulations are run for 2000 iterations using the SPSA tuner, with convergence generally beginning at around 1250 iterations. 
QISMET performs best, achieving a VQA expectation of -1.5, which is a 65\% improvement over the baseline.

The Blocking and Resampling schemes also show some improvement over the baseline.
While the Blocking scheme can avoid transient errors that move VQA away from the objective, it also hurts the ability to escape from local minima. 
Similarly, while the Resampling scheme's increase in samples offers some transient robustness, it has twice the computational cost per iteration and therefore can be 2x slower to converge.  
%TODO CAMERA
%Though these techniques are not specifically designing for transient noise, they both add some robustness against the fluctuating impact of transient noise.
%But these schemes perform the above techniques every iteration, meaning that robustness comes at the cost of slow convergence.
Therefore, the VQA improvements obtained from these are around 30\% lower than QISMET.
The 2nd-order scheme uses 2nd-order derivatives to influence the gradients calculated in each iteration.
We observe this scheme to in fact be detrimental in the presence of transients - imperfect 2nd-order derivatives can potentially skew the gradients even further away from optimal.
This scheme performs 35\% worse than the baseline and roughly 2.5x worse than QISMET.

\begin{figure}[t]
\centering
%\fbox{
\includegraphics[width=\columnwidth,trim={0cm 0cm 0cm 0cm},clip]{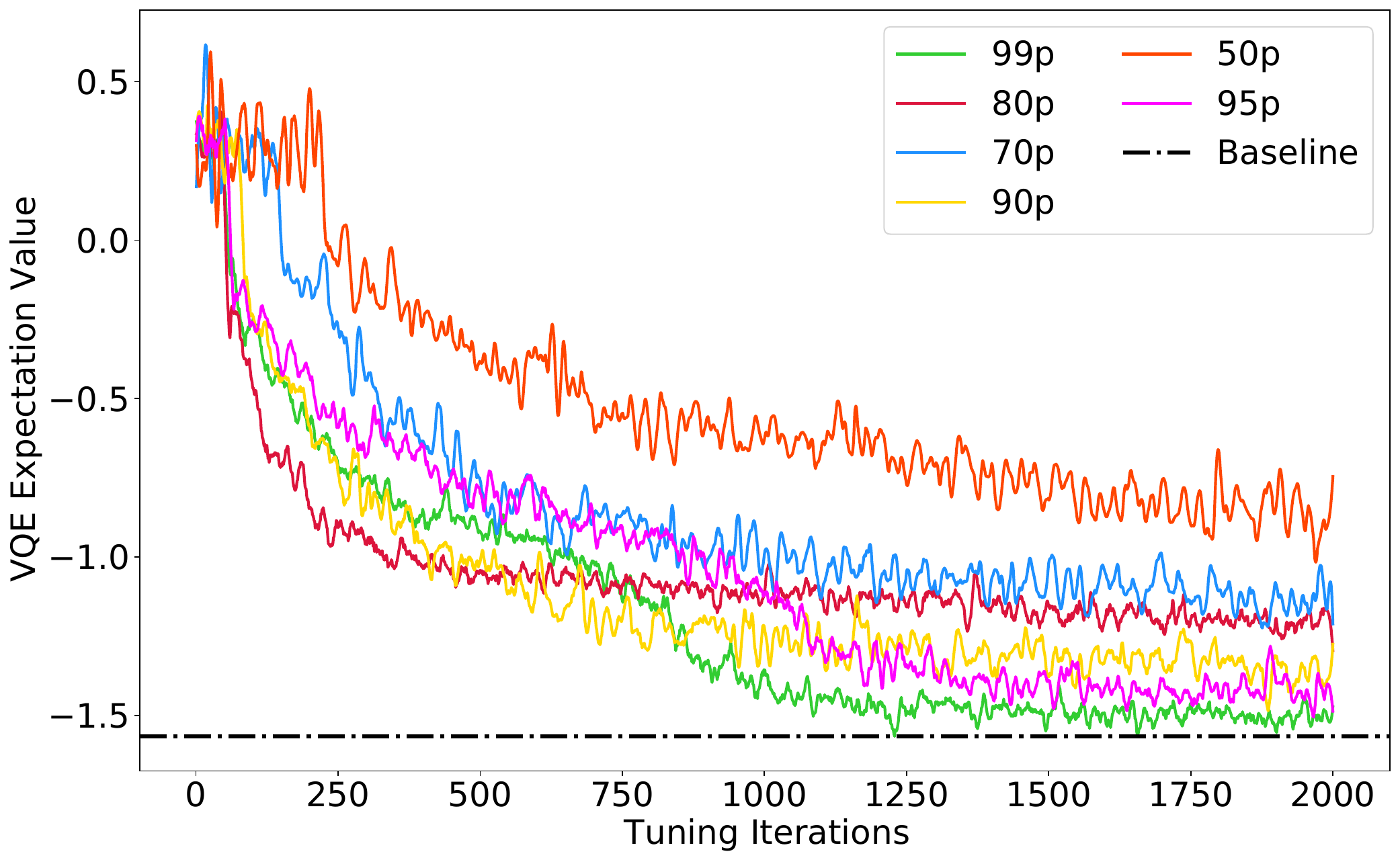}
%}
\caption{Only-transients skipping approach for App1. 
%Skipping thresholds vary from few skips (99p) to a high skip \% (50p). All versions perform worse than the baseline.
}
\label{fig:qismet_alt}
\end{figure}

\subsection{Alternative Only-Transients Skipping Approach}
\label{e_alt}

\iffalse
In the end of Section \ref{filter}, we noted that one could consider an alternative skipping approach which skips iterations simply based on the magnitude of the estimated transient error.
We argued such a solution is unsuitable because many transients that are constructive to VQA progress are skipped, affecting accuracy and substantially delaying convergence.
\fi

Fig.\ref{fig:qismet_alt} shows the the alternative skipping approach discussed in Section \ref{filter}, evaluated for App1.
Thresholds (based on which the controller decides to skip or not) vary from a very high threshold, which keeps the \% of skips below 1\% (99p), to a low threshold, which allows up to 50\% of the iterations to be skipped (50p). 
In all scenarios, the resulting VQA estimates are worse than the baseline and the higher skipping thresholds always perform better. 
This is clearly indicative that the skipping technique is more harmful than useful, motivating the need for a more intelligent approach such as QISMET.

\begin{figure}[t]
\centering
%\fbox{
\includegraphics[width=\columnwidth,trim={0cm 0cm 0cm 0cm},clip]{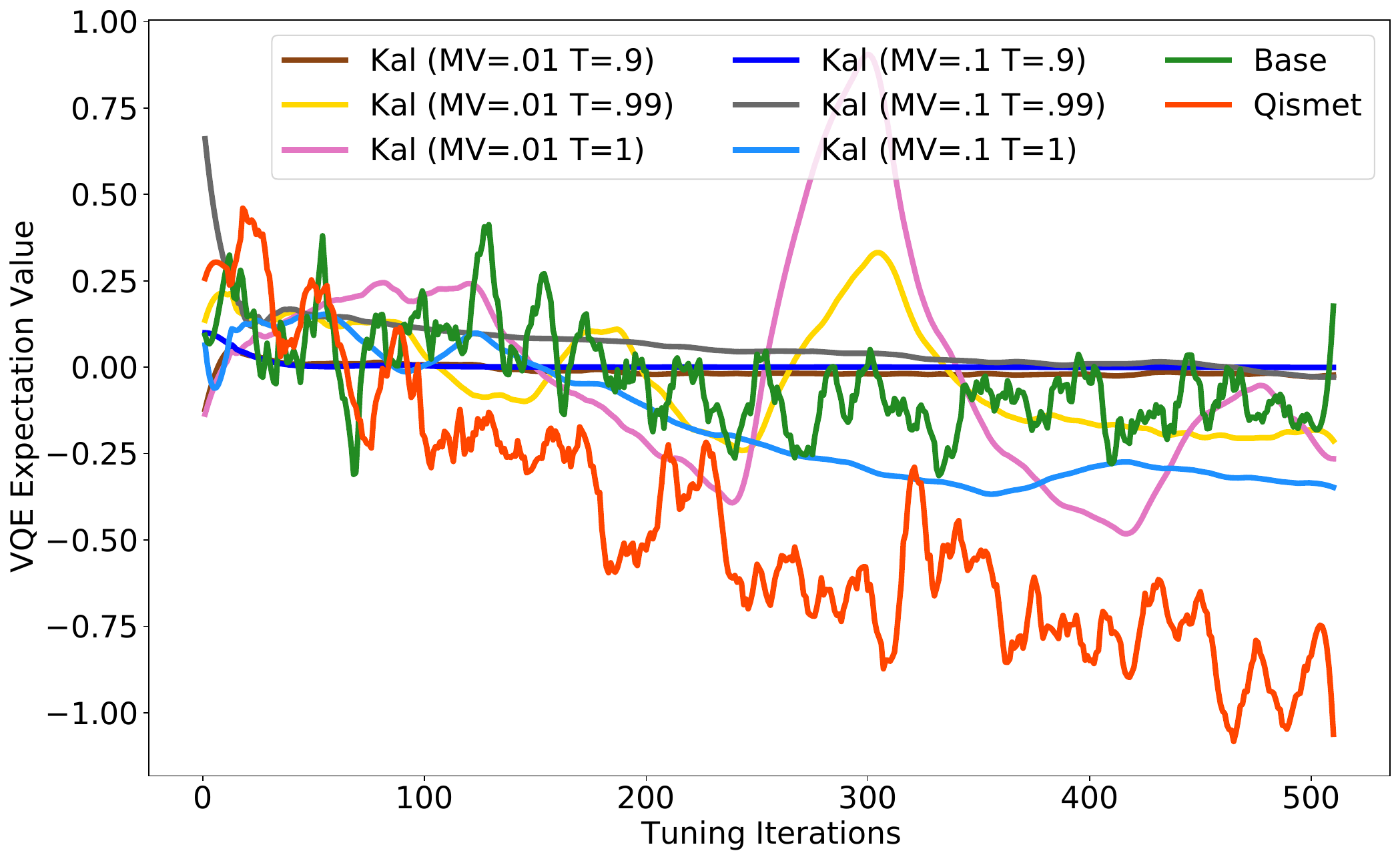}
%}
\caption{{Simulated comparison of Kalman filter against Qismet and baseline, for App6. Two Kalman hyper-parameters (described in Section \ref{eval_cmp}) are tuned over a limited search space. While good choices of Kalman hyper-parameters can provide some improvements over the baseline, the benefits are substantially lower than those from Qismet.}}
\label{fig:qismet_kal}
\end{figure}

\subsection{Comparing against Kalman filtering}
\label{kalman}
{
Furthering the discussion in Section \ref{filter}, we compare different Kalman instances (with different values of T and MV parameters - details in Section \ref{eval_cmp}) in Fig.\ref{fig:qismet_kal}.
We show its evaluation for App6, against the Baseline and Qismet, for 500 iterations, in simulation. 
Here, the Kalman filtering is applied on top of the noisy VQA tuning performed with SPSA. 
Note here that the T and MV values are tuned via an oracle approach, with noise known apriori, purely to maximize benefits. 
%In reality, the Kalman parameters would potentially have to be tuned over multiple end-to-end trials---but this is very expensive, apart from the fact that machine noise is highly unstable from one run to the next.
}

{We show 6 different Kalman instances with MV=0.01/0.1 and T=0.9/0.99/1 in Fig.\ref{fig:qismet_kal}.
A low MV value indicates that the filter assumes low influence of variation in the machine measurements - thus the filter does not actively combat it. 
Thus, any instance of high transient noise significantly impacts the filter (seen in the pink line, for example).
A high MV value indicates the possibility of high variation influence in the measurements - so the filter actively avoids this. 
Unfortunately, the filter is unable to differentiate between machine noise and actual algorithmic variance (navigating local minima etc) - thus, this filtering technique saturates quickly and poorly (seen in the dark blue line, for example).
A T value farther away from 1 imposes a forced downward descent on the estimated values - while this is helpful in true downward slopes, it is bad for escaping minima.
In this application example, the most stable low estimate is produced by (MV=0.1, T=1) whereas the lowest but an erratic estimate is produced by (MV=0.01, T=1).
While the estimates are as much as 1.4x better than the baseline, QISMET estimates are 3x better than the best Kalman variant, while also being stable.
}

\begin{figure}[t]
\centering
%\fbox{
\includegraphics[width=\columnwidth,trim={0cm 0cm 0cm 0cm},clip]{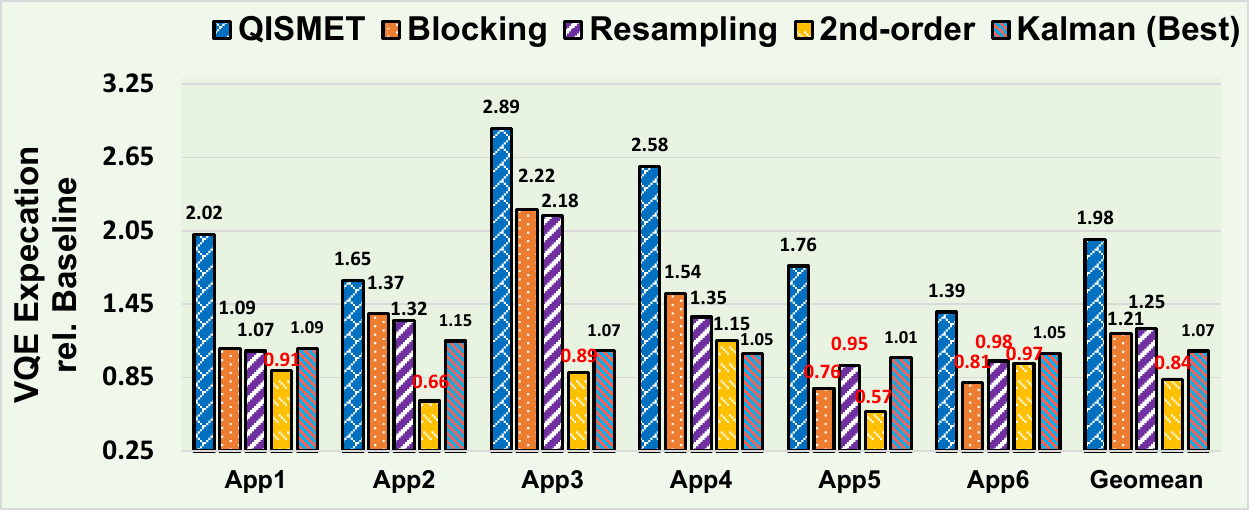}
%}
\caption{{QISMET benefits from the six simulated applications using the SPSA tuner run for 2000 iterations. QISMET provides consistent benefits,  outperforming the baseline and competing schemes (Blocking, Resampling, 2nd-order and Kalman filtering).} 
%The Blocking and Resampling schemes are inconsistent, offering mean improvement over the baseline but performing worse on some applications. 
%Greater QISMET benefits are obtained in the case of deeper circuits and/or on noisier machines.  
}
\label{fig:qismet_sim_full}
\end{figure}

\subsection{Simulated Comparisons for Multiple Benchmarks}
\label{e_sim}
Next, in Fig.\ref{fig:qismet_sim_full} we perform evaluation comparisons for five schemes - QISMET, Blocking, Resampling, 2nd-order {, and Kalman filter}, across the six applications listed in Table \ref{Table1}.
All evaluations are simulated for 2000 iterations and optimized with the SPSA tuner.
QISMET provides consistent benefits, always outperforming the baseline and competing schemes, achieving mean improvements of 2x (up to 3x), 1.7x, 1.6x, 2.4x, {and 1.85x} respectively. 
Note that Blocking and Resampling schemes are inconsistent, offering a mean improvement over the baseline (1.2-1.25x), but perform worse on some applications (highlighted in red). 
As noted earlier, while these schemes have some robustness to transients, they are slow to accurate convergence due to increased per-iteration evaluation.
%Further, being longer running schemes, they have a higher possibility of being hit by transients and it is intuitive that when the transients do have a detrimental impact (if a particularly tumultuous noisy phase occurs), these schemes can be hard hit. 
The 2nd-order scheme consistently performs worse than the baseline and is not observed to be a suitable solution for transient errors.
{The Kalman filter hyper-parameters are tuned as discussed in Section \ref{kalman}  and only the best-case results are shown. 
While there is non-negligible improvement over the baseline, the benefits are fairly low due to the reasons described in Section \ref{filter}.
Clearly, Qismet achieves a considerably higher improvement.
Also, the best Kalman instance varies across different applications, further highlighting the challenges.
}
Finally, it should be noted that QISMET benefits are generally greater for deeper circuits and/or on traces from noisier machines.
This is intuitive and in line with the discussion in Section \ref{m_circuit}.

\begin{figure}[h]
\centering
%\fbox{
\includegraphics[width=\columnwidth,trim={0cm 0cm 0cm 0cm},clip]{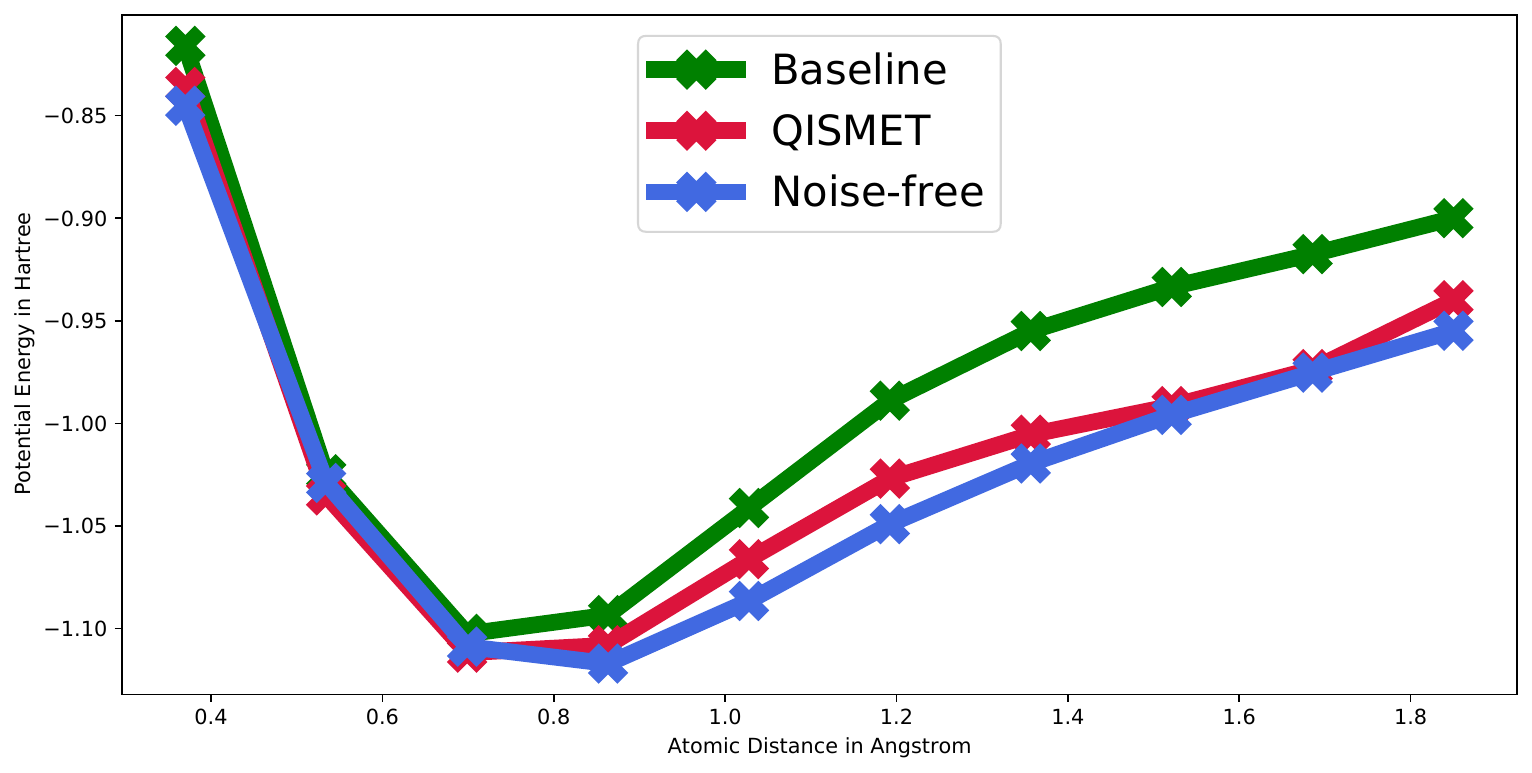}
%}
\caption{Effect of transient noise on multi-VQA experiments, shown by energy estimation of $H_2$. %Ten VQA instances are run at different $H-H$ bond lengths. QISMET produces high accuracy energy estimates, close to the noise-free scenario, whereas higher error is seen with the baseline.
}
\label{fig:qismet_sim_h2}
\end{figure}

\subsection{Multi-VQA Experiments for Molecules}
%Avoiding the detrimental effects of transients is of even greater importance in applications that require multiple segregated quantum experiments. %TODO CAMERA
In molecular chemistry, it is important to estimate the difference in VQE estimates across multiple Hamiltonians~\cite{vancoillie2016potential}.
Here, each Hamiltonian models a particular geometry of the molecule (usually different bond lengths), and the VQE estimate is a measure of the molecular potential energy.
The difference between the energy estimates at different bond lengths is indicative of the chemical reaction rates for the molecule of interest.
If transient noise affects some of the VQA experiments more than others, then the difference in energy estimates can be very skewed.

Fig.\ref{fig:qismet_sim_h2} shows potential energy estimation for the $H_2$ molecule through simulation.
This simulation setup only uses transient noise and no static noise component.
Estimations are performed for 10 different $H-H$ bond lengths, each corresponding to a unique Hamiltonian and VQE experiment.
The ideal noise-free scenario is shown, along with estimates obtained with QISMET and the baseline.
Clearly, QISMET provides high-accuracy estimates, and the shape of the bell curve closely models the noise-free scenario.
On the other hand, the baseline steadily deviates away from noise-free.
Greater deviation is expected at higher bond lengths since the quantum component of the potential energy is more significant in this region (and therefore, so is the quantum error).

\begin{figure}[h]
\centering
%\fbox{
\includegraphics[width=\columnwidth,trim={0cm 0cm 0cm 0cm},clip]{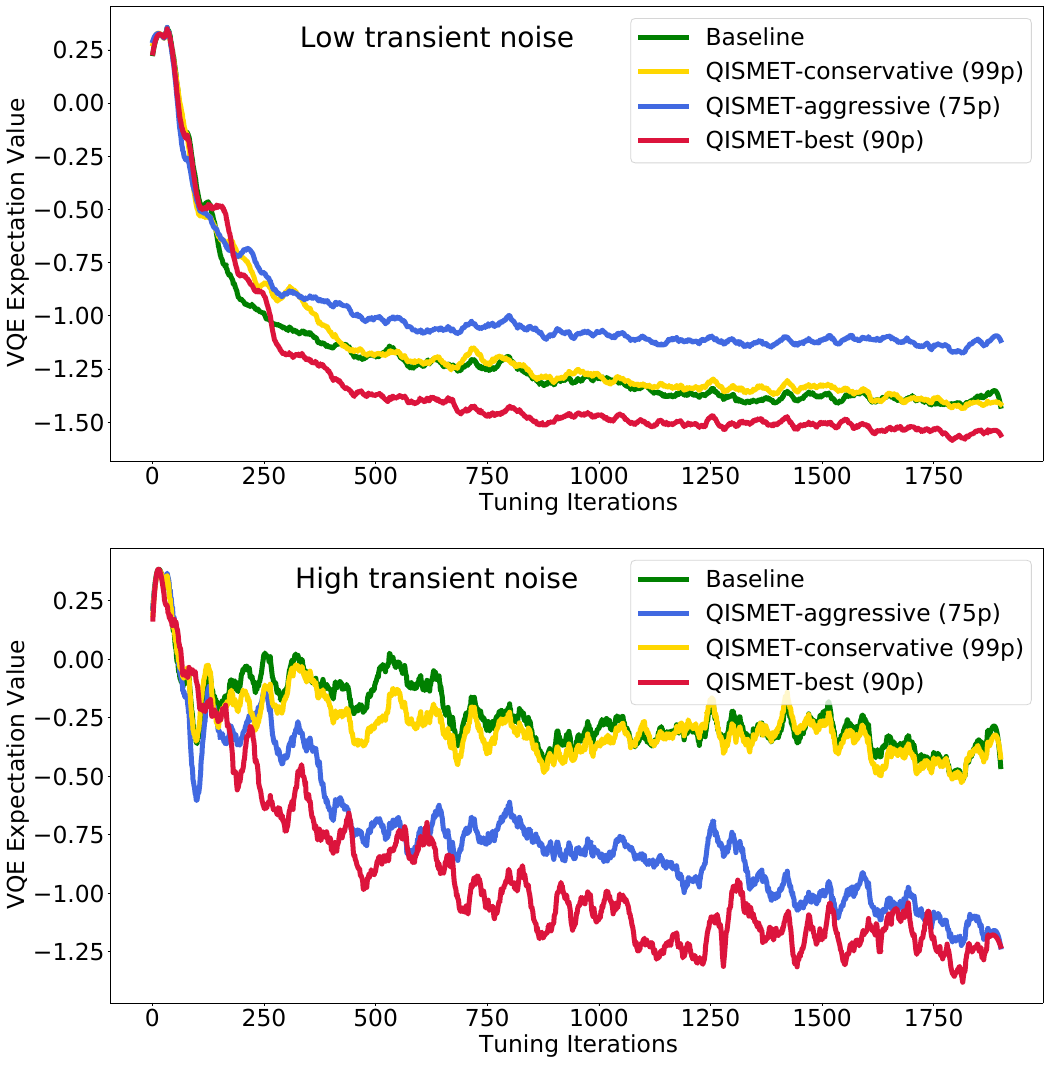}
%}
\caption{Different QISMET error thresholds are analyzed on two simulated use cases which experience low and high transient noise respectively. 
%The conservative threshold skips too few instances to have impact. The aggressive threshold pushes QISMET worse than the baseline when transients are low (too many unnecessary skips), but achieves considerable benefit in the high-transient error scenario (many impacting transients to skip). QISMET-best achieves the highest benefits in both scenarios.
}
\label{fig:qismet_sim_thresh}
\end{figure}

\subsection{ Sweeping the QISMET Error Threshold}
\label{e_thresh}
In Section \ref{d_controller}, we discussed the use of a threshold to control the fraction of transients that are skipped by the QISMET controller.
In Fig.\ref{fig:qismet_sim_thresh} three different thresholds are analyzed along with the baseline - these were introduced in Section \ref{6-method}.
The evaluation is shown for two simulated use cases which experience low and high transient noise respectively. 
For the chosen use cases, the conservative threshold skips too few instances to have a significant impact and, therefore, performs similarly to the baseline.
The aggressive threshold pushes QISMET to be worse than the baseline in the scenario with low transient noise - too many iterations with low transient noise are skipped unnecessarily which delay convergence. 
In the scenario with high transient noise, the aggressive threshold still achieves considerable benefit since there are enough such impacting transients to skip. 
The best-case threshold is a good trade-off and is able to achieve the highest benefits in both scenarios, achieving a 1.2x and 3x improvement over the baseline, respectively. 
Note that intelligent dynamic thresholding can potentially be used to improve these benefits further, but is beyond our current scope.

\section{Discussion}

\subsection{Tuning QISMET Parameters}
{There are two tunable parameters in the QISMET framework.
The more impactful one is the error threshold, which is discussed in Section \ref{d_controller} and evaluated in Section \ref{e_thresh}.
If transient errors are very few and of very large magnitude, then conservative thresholds are suitable---QISMET will predominantly follow the baseline and only avoid the transient errors which clearly stand out as egregious occurrences. 
On the other hand, transient errors which are more frequent and of magnitudes that are not too large but still detrimental, will benefit from more aggressive thresholds.}

{
The other tunable parameter is the retry budget or the maximum number of repetitions that QISMET will perform on any iteration that does not meet the error threshold. In this work we fix this to 5 repetitions. 
The optimal number for this budget could be derived from an understanding of the typical time duration for which transient effects last.
As long as the execution time taken by the retry-budget number of repetitions is slightly greater than the transient error duration, then an instance without the transient error's effects can be achieved for the given iteration.
At the same time, if some device characteristic change occurs for a long time duration, i.e., it is not transient (eg., machine re-calibration), then we would want the change to be quickly accepted by QISMET, and adapted to by the VQA tuner in future iterations. 
From our real device experiments, we observe that transient errors disappear in one or two repetitions. 
%We only observed the retry budget to actually be exhausted in around 10 instances across 100,000 possible instances.
}

\iffalse
{
%It is worth considering dynamic error thresholds and retry budgets based on heuristics or feedback based optimization schemes. 
Better understanding of device level characteristics and noise sources will be beneficial to tuning the QISMET parameters.
While the specifics of transient error characteristics are often unpredictable (such as the occurrence or magnitude of a TLS),
%due to limited understanding of noise in quantum devices (for example, predicting when a TLS or noise from classical control might occur is extremely challenging).
it could be possible that the probability of occurrence of impactful transient errors could be gauged from environmental deviations (like temperature), from features of the quantum circuit (like circuit depth), from qubit decoherence times etc. 
%On the one hand, this is a worthwhile future study, albeit requiring abundant experimentation and detailed information about device level characteristics.
%%such as temperature fluctuations on real quantum hardware. 
%%Such a detailed set of experiments is not feasible today due to limited time and access to commercial quantum devices, but  might be feasible in the future.%
%On the other hand, this is clear motivation for lightweight techniques like QISMET.
%are important today because they are able to effectively function in a lightweight manner without such detailed device knowledge.
}
\fi

\subsection{QISMET Suitability}
{
QISMET is most effective and has most significant benefits when transient errors are of high magnitude and of reasonably short duration.
Such instances can be identified and skipped in just a single trial.
Our empirical real machine observations suggest that this is often the case.
Transients of lower magnitude are not harmful to either QISMET or the baseline.
There are some particular scenarios in which QISMET could perform poorly.
One, if transient errors are gradually accumulating, then the errors would always be acceptable to the threshold but they might push VQA far from optimality.
In this scenario though, QISMET would perform no worse than the baseline.
Two, if high magnitude transient errors are of very long duration, then the effect of the transient is accepted after a retry budget worth of skips---thus, the negative impact on VQA estimation is similar to the baseline, but QISMET also suffers from lost iterations in this scenario---thus QISMET is worse off.
While there could be other adversarial scenarios that negatively impact QISMET, quantum devices are too noisy today for contrived adversarial patterns to have any realistic likelihood of occurrence over a prolonged period.
%And QISMET benefits on real quantum machines showcases this.
}

%rare and of considerable magnitude. 

\subsection{QISMET Overheads}
{
QISMET requires that each execution instance runs the current VQA iteration as well as a rerun of the previous iteration.
%rerun the circuit from the previous iteration. an additional time on each instance.
In the absence of other error mitigation and/or supporting circuits,  this means that the circuit execution overhead of QISMET is at least 2x (compared to a baseline with no transient errors).
For the above scenario, in the absence of no skips, the overhead is exactly 2x, whereas the overhead increases with the number of skips. 
On most fairly reliable machines, the number of skips is extremely small---only the egregious transients are being avoided. 
However, it should be noted that error mitigation circuits (such as those to combat measurement errors) are often run alongside the primary circuit.
These supporting circuits are often many in number, scaling in proportion to the number of measurement bits.
In the presence of such circuits, QISMET overheads are reduced.
Further, any QISMET overhead will only be negatively reflected if transients are entirely absent---in practical settings which do have some transients, avoiding them with QISMET substantially lowers the overall iteration and execution time cost.
%Finally, the impact of these QISMET overheads is rather minimal on execution time since the reference circuits are run in the same quantum `job' as the primary circuit. Intra-job overheads are considerably lower than inter-job overheads.
}

{
\textbf{Kalman comparison:} While, at first glance, it may seem that there are no overheads to the Kalman filtering (since circuits are run only once), it should be noted that the oracle (or any) approach used to tune the hyper-parameters requires some profiling, which would likely be much more expensive than the QISMET scenario, as described above.
}

\subsection{Other Classical Techniques}
{
In Sec.\ref{kalman}, we showed that QISMET can substantially outperform classical filtering techniques such as the Kalman filter.
Other filtering techniques are also available, such as Constant False Alarm Rate (CFAR)~\cite{CFAR} detection, which is used in radar systems to detect target returns against a background of noise and interference. 
Similar to Kalman, they are limited in their capability to avoid  specific detrimental transients.
%CFAR determines a power threshold above which any return can be considered to probably originate from a target, as opposed to a noise source. 
%Similar to the Kalman filter, efficient CFAR implementations have to rigorously analyze a wealth of radar data to create optimal threshold settings, which is problematic in the quantum world.
Finally, while the above filtering techniques are fairly unsupervised, supervised techniques are impractical since there is no useful labeled training data available, since the data would have to be  machine-application pair dependent and stable over long periods of time.}

\section{Conclusion}
\label{Conc}
Transient errors can be very detrimental to quantum computing in the NISQ era, especially for long-running applications such as VQAs.
\iffalse %TODO CAMERA
%VQA optimizers estimate gradients of some form~\cite{} across multiple execution instances, to iteratively choose the next set of VQA parameters.
%They work under the underlying assumption that the noise landscape of the device is unchanged during this gradient estimation process.
Noise fluctuation can cause a significant transient impact on the objective function estimation of some VQA instances, and this can severely affect gradients and therefore the accuracy and convergence of the VQA.
To our knowledge, this work is among the first to study the effects of transient errors on VQAs.

\fi
This paper proposes QISMET to navigate the dynamic noise landscape of VQA. 
It actively avoids instances of high fluctuating noise which can have a significant transient error impact on specific VQA iterations.
To achieve this, QISMET estimates transient noise in VQA iterations and designs a controller to keep the machine-obtained iteration gradients (specifically, their directions) faithful to their corresponding transient-free predicted estimates.
%By doing so, QISMET efficiently mitigates a major portion of the transient noise effects on VQAs.and is able to improve the fidelity of VQAs by 1.3x-3x over a traditional VQA baseline, across different applications and machines. %TODO CAMERA
 
%TODO CAMERA
%Further, to diligently analyze the effects of transients, this work also builds transient error noise models  from observing real device transients, which are then integrated with the Qiskit simulator.

%%%%%%%%%%%%%%%%%%%%%%%%%%

\begin{acks}
This work is funded in part by EPiQC, an NSF Expedition in Computing, under award CCF-1730449; 
in part by STAQ under award NSF Phy-1818914; in part by NSF award 2110860; 
in part by the US Department of Energy Office  of Advanced Scientific Computing Research, Accelerated Research for Quantum Computing Program; 
and in part by the NSF Quantum Leap Challenge Institute for Hybrid Quantum Architectures and Networks (NSF Award 2016136) 
and in part based upon work supported by the U.S. Department of Energy, Office of Science, National Quantum Information Science Research Centers.  
This work was completed in part with resources provided by
the University of Chicago's Research Computing Center.
This research used resources of the Oak Ridge Leadership Computing Facility, which is a DOE Office of Science User Facility supported under Contract DE-AC05-00OR22725.
GSR is supported as a Computing Innovation Fellow at the University of Chicago. This material is based upon work supported by the National Science Foundation under Grant \# 2030859 to the Computing Research Association for the CIFellows Project.
KNS is supported by IBM as a Postdoctoral Scholar at the University of Chicago and the Chicago Quantum Exchange.
HH is supported by NSF (grants CCF-2119184, CNS-1956180, CNS-1952050, CCF-1823032, CNS-1764039), ARO (grant W911NF1920321), and a DOE Early Career Award (grant DESC0014195 0003).
FTC is Chief Scientist for Quantum Software at ColdQuanta and an advisor to Quantum Circuits, Inc.
\end{acks}

%%
%% The next two lines define the bibliography style to be used, and
%% the bibliography file.
\balance
\bibliographystyle{ACM-Reference-Format}
\bibliography{refs}

%%% -*-BibTeX-*-
%%% Do NOT edit. File created by BibTeX with style
%%% ACM-Reference-Format-Journals [18-Jan-2012].

\begin{thebibliography}{45}

%%% ====================================================================
%%% NOTE TO THE USER: you can override these defaults by providing
%%% customized versions of any of these macros before the \bibliography
%%% command.  Each of them MUST provide its own final punctuation,
%%% except for \shownote{}, \showDOI{}, and \showURL{}.  The latter two
%%% do not use final punctuation, in order to avoid confusing it with
%%% the Web address.
%%%
%%% To suppress output of a particular field, define its macro to expand
%%% to an empty string, or better, \unskip, like this:
%%%
%%% \newcommand{\showDOI}[1]{\unskip}   % LaTeX syntax
%%%
%%% \def \showDOI #1{\unskip}           % plain TeX syntax
%%%
%%% ====================================================================

\ifx \showCODEN    \undefined \def \showCODEN     #1{\unskip}     \fi
\ifx \showDOI      \undefined \def \showDOI       #1{#1}\fi
\ifx \showISBNx    \undefined \def \showISBNx     #1{\unskip}     \fi
\ifx \showISBNxiii \undefined \def \showISBNxiii  #1{\unskip}     \fi
\ifx \showISSN     \undefined \def \showISSN      #1{\unskip}     \fi
\ifx \showLCCN     \undefined \def \showLCCN      #1{\unskip}     \fi
\ifx \shownote     \undefined \def \shownote      #1{#1}          \fi
\ifx \showarticletitle \undefined \def \showarticletitle #1{#1}   \fi
\ifx \showURL      \undefined \def \showURL       {\relax}        \fi
% The following commands are used for tagged output and should be
% invisible to TeX
\providecommand\bibfield[2]{#2}
\providecommand\bibinfo[2]{#2}
\providecommand\natexlab[1]{#1}
\providecommand\showeprint[2][]{arXiv:#2}

\bibitem[IBM(2021a)]%
        {IBM-RA}
 \bibinfo{year}{2021}\natexlab{a}.
\newblock \bibinfo{title}{IBM Quantum RA ansatz}.
\newblock
  \bibinfo{howpublished}{\url{https://qiskit.org/documentation/stubs/qiskit.circuit.library.RealAmplitudes.html}}.
\newblock


\bibitem[IBM(2021b)]%
        {IBM-SU2}
 \bibinfo{year}{2021}\natexlab{b}.
\newblock \bibinfo{title}{IBM Quantum SU2 ansatz}.
\newblock
  \bibinfo{howpublished}{\url{https://qiskit.org/documentation/stubs/qiskit.circuit.library.EfficientSU2.html}}.
\newblock


\bibitem[IBM(2021c)]%
        {IBMQS}
 \bibinfo{year}{2021}\natexlab{c}.
\newblock \bibinfo{title}{IBM Quantum Systems}.
\newblock
  \bibinfo{howpublished}{\url{https://quantum-computing.ibm.com/services?systems=all}}.
\newblock


\bibitem[SPS(2021)]%
        {SPSA}
 \bibinfo{year}{2021}\natexlab{}.
\newblock \bibinfo{title}{SPSA: Simultaneous Perturbation Stochastic
  Approximation method}.
\newblock \bibinfo{howpublished}{\url{https://www.jhuapl.edu/spsa/ }}.
\newblock


\bibitem[Abraham et~al\mbox{.}(2019)]%
        {Qiskit}
\bibfield{author}{\bibinfo{person}{H{\'e}ctor Abraham},
  \bibinfo{person}{AduOffei}, \bibinfo{person}{Rochisha Agarwal},
  \bibinfo{person}{Ismail~Yunus Akhalwaya}, \bibinfo{person}{Gadi
  Aleksandrowicz}, \bibinfo{person}{Thomas Alexander}, \bibinfo{person}{Matthew
  Amy}, \bibinfo{person}{Eli Arbel}, \bibinfo{person}{Arijit02},
  \bibinfo{person}{Abraham Asfaw}, \bibinfo{person}{Artur Avkhadiev},
  \bibinfo{person}{Carlos Azaustre}, \bibinfo{person}{AzizNgoueya},
  \bibinfo{person}{Abhik Banerjee}, \bibinfo{person}{Aman Bansal},
  \bibinfo{person}{Panagiotis Barkoutsos}, \bibinfo{person}{George Barron},
  \bibinfo{person}{George~S. Barron}, \bibinfo{person}{Luciano Bello},
  \bibinfo{person}{Yael Ben-Haim}, \bibinfo{person}{Daniel Bevenius},
  \bibinfo{person}{Arjun Bhobe}, \bibinfo{person}{Lev~S. Bishop},
  \bibinfo{person}{Carsten Blank}, \bibinfo{person}{Sorin Bolos},
  \bibinfo{person}{Samuel Bosch}, \bibinfo{person}{Brandon},
  \bibinfo{person}{Sergey Bravyi}, \bibinfo{person}{Bryce-Fuller},
  \bibinfo{person}{David Bucher}, \bibinfo{person}{Artemiy Burov},
  \bibinfo{person}{Fran Cabrera}, \bibinfo{person}{Padraic Calpin},
  \bibinfo{person}{Lauren Capelluto}, \bibinfo{person}{Jorge Carballo},
  \bibinfo{person}{Gin{\'e}s Carrascal}, \bibinfo{person}{Adrian Chen},
  \bibinfo{person}{Chun-Fu Chen}, \bibinfo{person}{Edward Chen},
  \bibinfo{person}{Jielun~(Chris) Chen}, \bibinfo{person}{Richard Chen},
  \bibinfo{person}{Jerry~M. Chow}, \bibinfo{person}{Spencer Churchill},
  \bibinfo{person}{Christian Claus}, \bibinfo{person}{Christian Clauss},
  \bibinfo{person}{Romilly Cocking}, \bibinfo{person}{Filipe Correa},
  \bibinfo{person}{Abigail~J. Cross}, \bibinfo{person}{Andrew~W. Cross},
  \bibinfo{person}{Simon Cross}, \bibinfo{person}{Juan Cruz-Benito},
  \bibinfo{person}{Chris Culver}, \bibinfo{person}{Antonio~D.
  C{\'o}rcoles-Gonzales}, \bibinfo{person}{Sean Dague},
  \bibinfo{person}{Tareq~El Dandachi}, \bibinfo{person}{Marcus Daniels},
  \bibinfo{person}{Matthieu Dartiailh}, \bibinfo{person}{DavideFrr},
  \bibinfo{person}{Abd{\'o}n~Rodr{\'\i}guez Davila}, \bibinfo{person}{Anton
  Dekusar}, \bibinfo{person}{Delton Ding}, \bibinfo{person}{Jun Doi},
  \bibinfo{person}{Eric Drechsler}, \bibinfo{person}{Drew},
  \bibinfo{person}{Eugene Dumitrescu}, \bibinfo{person}{Karel Dumon},
  \bibinfo{person}{Ivan Duran}, \bibinfo{person}{Kareem EL-Safty},
  \bibinfo{person}{Eric Eastman}, \bibinfo{person}{Grant Eberle},
  \bibinfo{person}{Pieter Eendebak}, \bibinfo{person}{Daniel Egger},
  \bibinfo{person}{Mark Everitt}, \bibinfo{person}{Paco~Mart{\'\i}n
  Fern{\'a}ndez}, \bibinfo{person}{Axel~Hern{\'a}ndez Ferrera},
  \bibinfo{person}{Romain Fouilland}, \bibinfo{person}{FranckChevallier},
  \bibinfo{person}{Albert Frisch}, \bibinfo{person}{Andreas Fuhrer},
  \bibinfo{person}{Bryce Fuller}, \bibinfo{person}{MELVIN GEORGE},
  \bibinfo{person}{Julien Gacon}, \bibinfo{person}{Borja~Godoy Gago},
  \bibinfo{person}{Claudio Gambella}, \bibinfo{person}{Jay~M. Gambetta},
  \bibinfo{person}{Adhisha Gammanpila}, \bibinfo{person}{Luis Garcia},
  \bibinfo{person}{Tanya Garg}, \bibinfo{person}{Shelly Garion},
  \bibinfo{person}{Austin Gilliam}, \bibinfo{person}{Aditya Giridharan},
  \bibinfo{person}{Juan Gomez-Mosquera}, \bibinfo{person}{Salvador de~la
  Puente~Gonz{\'a}lez}, \bibinfo{person}{Jesse Gorzinski}, \bibinfo{person}{Ian
  Gould}, \bibinfo{person}{Donny Greenberg}, \bibinfo{person}{Dmitry Grinko},
  \bibinfo{person}{Wen Guan}, \bibinfo{person}{John~A. Gunnels},
  \bibinfo{person}{Mikael Haglund}, \bibinfo{person}{Isabel Haide},
  \bibinfo{person}{Ikko Hamamura}, \bibinfo{person}{Omar~Costa Hamido},
  \bibinfo{person}{Frank Harkins}, \bibinfo{person}{Vojtech Havlicek},
  \bibinfo{person}{Joe Hellmers}, \bibinfo{person}{{\L}ukasz Herok},
  \bibinfo{person}{Stefan Hillmich}, \bibinfo{person}{Hiroshi Horii},
  \bibinfo{person}{Connor Howington}, \bibinfo{person}{Shaohan Hu},
  \bibinfo{person}{Wei Hu}, \bibinfo{person}{Junye Huang},
  \bibinfo{person}{Rolf Huisman}, \bibinfo{person}{Haruki Imai},
  \bibinfo{person}{Takashi Imamichi}, \bibinfo{person}{Kazuaki Ishizaki},
  \bibinfo{person}{Raban Iten}, \bibinfo{person}{Toshinari Itoko},
  \bibinfo{person}{JamesSeaward}, \bibinfo{person}{Ali Javadi},
  \bibinfo{person}{Ali Javadi-Abhari}, \bibinfo{person}{Jessica},
  \bibinfo{person}{Madhav Jivrajani}, \bibinfo{person}{Kiran Johns},
  \bibinfo{person}{Scott Johnstun}, \bibinfo{person}{Jonathan-Shoemaker},
  \bibinfo{person}{Vismai K}, \bibinfo{person}{Tal Kachmann},
  \bibinfo{person}{Naoki Kanazawa}, \bibinfo{person}{Kang-Bae},
  \bibinfo{person}{Anton Karazeev}, \bibinfo{person}{Paul Kassebaum},
  \bibinfo{person}{Josh Kelso}, \bibinfo{person}{Spencer King},
  \bibinfo{person}{Knabberjoe}, \bibinfo{person}{Yuri Kobayashi},
  \bibinfo{person}{Arseny Kovyrshin}, \bibinfo{person}{Rajiv Krishnakumar},
  \bibinfo{person}{Vivek Krishnan}, \bibinfo{person}{Kevin Krsulich},
  \bibinfo{person}{Prasad Kumkar}, \bibinfo{person}{Gawel Kus},
  \bibinfo{person}{Ryan LaRose}, \bibinfo{person}{Enrique Lacal},
  \bibinfo{person}{Rapha{\"e}l Lambert}, \bibinfo{person}{John Lapeyre},
  \bibinfo{person}{Joe Latone}, \bibinfo{person}{Scott Lawrence},
  \bibinfo{person}{Christina Lee}, \bibinfo{person}{Gushu Li},
  \bibinfo{person}{Dennis Liu}, \bibinfo{person}{Peng Liu},
  \bibinfo{person}{Yunho Maeng}, \bibinfo{person}{Kahan Majmudar},
  \bibinfo{person}{Aleksei Malyshev}, \bibinfo{person}{Joshua Manela},
  \bibinfo{person}{Jakub Marecek}, \bibinfo{person}{Manoel Marques},
  \bibinfo{person}{Dmitri Maslov}, \bibinfo{person}{Dolph Mathews},
  \bibinfo{person}{Atsushi Matsuo}, \bibinfo{person}{Douglas~T. McClure},
  \bibinfo{person}{Cameron McGarry}, \bibinfo{person}{David McKay},
  \bibinfo{person}{Dan McPherson}, \bibinfo{person}{Srujan Meesala},
  \bibinfo{person}{Thomas Metcalfe}, \bibinfo{person}{Martin Mevissen},
  \bibinfo{person}{Andrew Meyer}, \bibinfo{person}{Antonio Mezzacapo},
  \bibinfo{person}{Rohit Midha}, \bibinfo{person}{Zlatko Minev},
  \bibinfo{person}{Abby Mitchell}, \bibinfo{person}{Nikolaj Moll},
  \bibinfo{person}{Jhon Montanez}, \bibinfo{person}{Michael~Duane Mooring},
  \bibinfo{person}{Renier Morales}, \bibinfo{person}{Niall Moran},
  \bibinfo{person}{Mario Motta}, \bibinfo{person}{MrF},
  \bibinfo{person}{Prakash Murali}, \bibinfo{person}{Jan M{\"u}ggenburg},
  \bibinfo{person}{David Nadlinger}, \bibinfo{person}{Ken Nakanishi},
  \bibinfo{person}{Giacomo Nannicini}, \bibinfo{person}{Paul Nation},
  \bibinfo{person}{Edwin Navarro}, \bibinfo{person}{Yehuda Naveh},
  \bibinfo{person}{Scott~Wyman Neagle}, \bibinfo{person}{Patrick Neuweiler},
  \bibinfo{person}{Johan Nicander}, \bibinfo{person}{Pradeep Niroula},
  \bibinfo{person}{Hassi Norlen}, \bibinfo{person}{NuoWenLei},
  \bibinfo{person}{Lee~James O'Riordan}, \bibinfo{person}{Oluwatobi Ogunbayo},
  \bibinfo{person}{Pauline Ollitrault}, \bibinfo{person}{Raul Otaolea},
  \bibinfo{person}{Steven Oud}, \bibinfo{person}{Dan Padilha},
  \bibinfo{person}{Hanhee Paik}, \bibinfo{person}{Soham Pal},
  \bibinfo{person}{Yuchen Pang}, \bibinfo{person}{Simone Perriello},
  \bibinfo{person}{Anna Phan}, \bibinfo{person}{Francesco Piro},
  \bibinfo{person}{Marco Pistoia}, \bibinfo{person}{Christophe Piveteau},
  \bibinfo{person}{Pierre Pocreau}, \bibinfo{person}{Alejandro
  Pozas-iKerstjens}, \bibinfo{person}{Viktor Prutyanov},
  \bibinfo{person}{Daniel Puzzuoli}, \bibinfo{person}{Jes{\'u}s P{\'e}rez},
  \bibinfo{person}{Quintiii}, \bibinfo{person}{Rafey~Iqbal Rahman},
  \bibinfo{person}{Arun Raja}, \bibinfo{person}{Nipun Ramagiri},
  \bibinfo{person}{Anirudh Rao}, \bibinfo{person}{Rudy Raymond},
  \bibinfo{person}{Rafael Mart{\'\i}n-Cuevas Redondo}, \bibinfo{person}{Max
  Reuter}, \bibinfo{person}{Julia Rice}, \bibinfo{person}{Marcello~La Rocca},
  \bibinfo{person}{Diego~M. Rodr{\'\i}guez}, \bibinfo{person}{RohithKarur},
  \bibinfo{person}{Max Rossmannek}, \bibinfo{person}{Mingi Ryu},
  \bibinfo{person}{Tharrmashastha SAPV}, \bibinfo{person}{SamFerracin},
  \bibinfo{person}{Martin Sandberg}, \bibinfo{person}{Hirmay Sandesara},
  \bibinfo{person}{Ritvik Sapra}, \bibinfo{person}{Hayk Sargsyan},
  \bibinfo{person}{Aniruddha Sarkar}, \bibinfo{person}{Ninad Sathaye},
  \bibinfo{person}{Bruno Schmitt}, \bibinfo{person}{Chris Schnabel},
  \bibinfo{person}{Zachary Schoenfeld}, \bibinfo{person}{Travis~L. Scholten},
  \bibinfo{person}{Eddie Schoute}, \bibinfo{person}{Joachim Schwarm},
  \bibinfo{person}{Ismael~Faro Sertage}, \bibinfo{person}{Kanav Setia},
  \bibinfo{person}{Nathan Shammah}, \bibinfo{person}{Yunong Shi},
  \bibinfo{person}{Adenilton Silva}, \bibinfo{person}{Andrea Simonetto},
  \bibinfo{person}{Nick Singstock}, \bibinfo{person}{Yukio Siraichi},
  \bibinfo{person}{Iskandar Sitdikov}, \bibinfo{person}{Seyon Sivarajah},
  \bibinfo{person}{Magnus~Berg Sletfjerding}, \bibinfo{person}{John~A. Smolin},
  \bibinfo{person}{Mathias Soeken}, \bibinfo{person}{Igor~Olegovich Sokolov},
  \bibinfo{person}{Igor Sokolov}, \bibinfo{person}{SooluThomas},
  \bibinfo{person}{Starfish}, \bibinfo{person}{Dominik Steenken},
  \bibinfo{person}{Matt Stypulkoski}, \bibinfo{person}{Shaojun Sun},
  \bibinfo{person}{Kevin~J. Sung}, \bibinfo{person}{Hitomi Takahashi},
  \bibinfo{person}{Tanvesh Takawale}, \bibinfo{person}{Ivano Tavernelli},
  \bibinfo{person}{Charles Taylor}, \bibinfo{person}{Pete Taylour},
  \bibinfo{person}{Soolu Thomas}, \bibinfo{person}{Mathieu Tillet},
  \bibinfo{person}{Maddy Tod}, \bibinfo{person}{Miroslav Tomasik},
  \bibinfo{person}{Enrique de~la Torre}, \bibinfo{person}{Kenso Trabing},
  \bibinfo{person}{Matthew Treinish}, \bibinfo{person}{TrishaPe},
  \bibinfo{person}{Davindra Tulsi}, \bibinfo{person}{Wes Turner},
  \bibinfo{person}{Yotam Vaknin}, \bibinfo{person}{Carmen~Recio Valcarce},
  \bibinfo{person}{Francois Varchon}, \bibinfo{person}{Almudena~Carrera
  Vazquez}, \bibinfo{person}{Victor Villar}, \bibinfo{person}{Desiree
  Vogt-Lee}, \bibinfo{person}{Christophe Vuillot}, \bibinfo{person}{James
  Weaver}, \bibinfo{person}{Johannes Weidenfeller}, \bibinfo{person}{Rafal
  Wieczorek}, \bibinfo{person}{Jonathan~A. Wildstrom}, \bibinfo{person}{Erick
  Winston}, \bibinfo{person}{Jack~J. Woehr}, \bibinfo{person}{Stefan Woerner},
  \bibinfo{person}{Ryan Woo}, \bibinfo{person}{Christopher~J. Wood},
  \bibinfo{person}{Ryan Wood}, \bibinfo{person}{Stephen Wood},
  \bibinfo{person}{Steve Wood}, \bibinfo{person}{James Wootton},
  \bibinfo{person}{Daniyar Yeralin}, \bibinfo{person}{David Yonge-Mallo},
  \bibinfo{person}{Richard Young}, \bibinfo{person}{Jessie Yu},
  \bibinfo{person}{Christopher Zachow}, \bibinfo{person}{Laura Zdanski},
  \bibinfo{person}{Helena Zhang}, \bibinfo{person}{Christa Zoufal},
  \bibinfo{person}{Zoufalc}, \bibinfo{person}{a kapila}, \bibinfo{person}{a
  matsuo}, \bibinfo{person}{bcamorrison}, \bibinfo{person}{brandhsn},
  \bibinfo{person}{nick bronn}, \bibinfo{person}{chlorophyll zz},
  \bibinfo{person}{dekel.meirom}, \bibinfo{person}{dekelmeirom},
  \bibinfo{person}{dekool}, \bibinfo{person}{dime10},
  \bibinfo{person}{drholmie}, \bibinfo{person}{dtrenev},
  \bibinfo{person}{ehchen}, \bibinfo{person}{elfrocampeador},
  \bibinfo{person}{faisaldebouni}, \bibinfo{person}{fanizzamarco},
  \bibinfo{person}{gabrieleagl}, \bibinfo{person}{gadial},
  \bibinfo{person}{galeinston}, \bibinfo{person}{georgios ts},
  \bibinfo{person}{gruu}, \bibinfo{person}{hhorii},
  \bibinfo{person}{hykavitha}, \bibinfo{person}{jagunther},
  \bibinfo{person}{jliu45}, \bibinfo{person}{jscott2},
  \bibinfo{person}{kanejess}, \bibinfo{person}{klinvill},
  \bibinfo{person}{krutik2966}, \bibinfo{person}{kurarrr},
  \bibinfo{person}{lerongil}, \bibinfo{person}{ma5x}, \bibinfo{person}{merav
  aharoni}, \bibinfo{person}{michelle4654}, \bibinfo{person}{ordmoj},
  \bibinfo{person}{sagar pahwa}, \bibinfo{person}{rmoyard},
  \bibinfo{person}{saswati qiskit}, \bibinfo{person}{scottkelso},
  \bibinfo{person}{sethmerkel}, \bibinfo{person}{strickroman},
  \bibinfo{person}{sumitpuri}, \bibinfo{person}{tigerjack},
  \bibinfo{person}{toural}, \bibinfo{person}{tsura crisaldo},
  \bibinfo{person}{vvilpas}, \bibinfo{person}{welien},
  \bibinfo{person}{willhbang}, \bibinfo{person}{yang.luh},
  \bibinfo{person}{yotamvakninibm}, {and} \bibinfo{person}{Mantas
  {\v{C}}epulkovskis}.} \bibinfo{year}{2019}\natexlab{}.
\newblock \bibinfo{title}{Qiskit: An Open-source Framework for Quantum
  Computing}.
\newblock
\newblock
\urldef\tempurl%
\url{https://doi.org/10.5281/zenodo.2562110}
\showDOI{\tempurl}


\bibitem[Biamonte et~al\mbox{.}(2017)]%
        {biamonte2017quantum}
\bibfield{author}{\bibinfo{person}{Jacob Biamonte}, \bibinfo{person}{Peter
  Wittek}, \bibinfo{person}{Nicola Pancotti}, \bibinfo{person}{Patrick
  Rebentrost}, \bibinfo{person}{Nathan Wiebe}, {and} \bibinfo{person}{Seth
  Lloyd}.} \bibinfo{year}{2017}\natexlab{}.
\newblock \showarticletitle{Quantum machine learning}.
\newblock \bibinfo{journal}{\emph{Nature}} \bibinfo{volume}{549},
  \bibinfo{number}{7671} (\bibinfo{year}{2017}), \bibinfo{pages}{195--202}.
\newblock


\bibitem[Biercuk et~al\mbox{.}(2011)]%
        {DDBiercuk_2011}
\bibfield{author}{\bibinfo{person}{M~J Biercuk}, \bibinfo{person}{A~C Doherty},
  {and} \bibinfo{person}{H Uys}.} \bibinfo{year}{2011}\natexlab{}.
\newblock \showarticletitle{Dynamical decoupling sequence construction as a
  filter-design problem}.
\newblock \bibinfo{journal}{\emph{Journal of Physics B: Atomic, Molecular and
  Optical Physics}} \bibinfo{volume}{44}, \bibinfo{number}{15}
  (\bibinfo{date}{Jul} \bibinfo{year}{2011}), \bibinfo{pages}{154002}.
\newblock
\showISSN{1361-6455}
\urldef\tempurl%
\url{https://doi.org/10.1088/0953-4075/44/15/154002}
\showDOI{\tempurl}


\bibitem[Bravyi et~al\mbox{.}(2021)]%
        {bravyi2021mitigating}
\bibfield{author}{\bibinfo{person}{Sergey Bravyi}, \bibinfo{person}{Sarah
  Sheldon}, \bibinfo{person}{Abhinav Kandala}, \bibinfo{person}{David~C Mckay},
  {and} \bibinfo{person}{Jay~M Gambetta}.} \bibinfo{year}{2021}\natexlab{}.
\newblock \showarticletitle{Mitigating measurement errors in multiqubit
  experiments}.
\newblock \bibinfo{journal}{\emph{Physical Review A}} \bibinfo{volume}{103},
  \bibinfo{number}{4} (\bibinfo{year}{2021}), \bibinfo{pages}{042605}.
\newblock


\bibitem[Burnett et~al\mbox{.}(2019)]%
        {burnett2019decoherence}
\bibfield{author}{\bibinfo{person}{JJ Burnett}, \bibinfo{person}{A Bengtsson},
  \bibinfo{person}{M Scigliuzzo}, \bibinfo{person}{D Niepce},
  \bibinfo{person}{M Kudra}, \bibinfo{person}{P Delsing}, {and}
  \bibinfo{person}{J Bylander}.} \bibinfo{year}{2019}\natexlab{}.
\newblock \showarticletitle{Decoherence benchmarking of superconducting qubits.
  npj Quantum Inf. 5}.
\newblock  (\bibinfo{year}{2019}).
\newblock


\bibitem[Ding et~al\mbox{.}(2020)]%
        {ding2020systematic}
\bibfield{author}{\bibinfo{person}{Yongshan Ding}, \bibinfo{person}{Pranav
  Gokhale}, \bibinfo{person}{Sophia~Fuhui Lin}, \bibinfo{person}{Richard
  Rines}, \bibinfo{person}{Thomas Propson}, {and} \bibinfo{person}{Frederic~T
  Chong}.} \bibinfo{year}{2020}\natexlab{}.
\newblock \showarticletitle{Systematic Crosstalk Mitigation for Superconducting
  Qubits via Frequency-Aware Compilation}.
\newblock \bibinfo{journal}{\emph{arXiv preprint arXiv:2008.09503}}
  (\bibinfo{year}{2020}).
\newblock


\bibitem[Farhi et~al\mbox{.}(2014)]%
        {farhi2014quantum}
\bibfield{author}{\bibinfo{person}{Edward Farhi}, \bibinfo{person}{Jeffrey
  Goldstone}, {and} \bibinfo{person}{Sam Gutmann}.}
  \bibinfo{year}{2014}\natexlab{}.
\newblock \bibinfo{title}{A Quantum Approximate Optimization Algorithm}.
\newblock
\newblock
\showeprint[arxiv]{1411.4028}~[quant-ph]


\bibitem[Gambetta et~al\mbox{.}(2017)]%
        {gambetta2017building}
\bibfield{author}{\bibinfo{person}{Jay~M Gambetta}, \bibinfo{person}{Jerry~M
  Chow}, {and} \bibinfo{person}{Matthias Steffen}.}
  \bibinfo{year}{2017}\natexlab{}.
\newblock \showarticletitle{Building logical qubits in a superconducting
  quantum computing system}.
\newblock \bibinfo{journal}{\emph{npj Quantum Information}}
  \bibinfo{volume}{3}, \bibinfo{number}{1} (\bibinfo{year}{2017}),
  \bibinfo{pages}{1--7}.
\newblock


\bibitem[Giurgica-Tiron et~al\mbox{.}(2020)]%
        {giurgica2020digital}
\bibfield{author}{\bibinfo{person}{Tudor Giurgica-Tiron},
  \bibinfo{person}{Yousef Hindy}, \bibinfo{person}{Ryan LaRose},
  \bibinfo{person}{Andrea Mari}, {and} \bibinfo{person}{William~J Zeng}.}
  \bibinfo{year}{2020}\natexlab{}.
\newblock \showarticletitle{Digital zero noise extrapolation for quantum error
  mitigation}. In \bibinfo{booktitle}{\emph{2020 IEEE International Conference
  on Quantum Computing and Engineering (QCE)}}. IEEE,
  \bibinfo{pages}{306--316}.
\newblock


\bibitem[Gokhale et~al\mbox{.}(2019)]%
        {Gokhale:2019}
\bibfield{author}{\bibinfo{person}{Pranav Gokhale}, \bibinfo{person}{Yongshan
  Ding}, \bibinfo{person}{Thomas Propson}, \bibinfo{person}{Christopher
  Winkler}, \bibinfo{person}{Nelson Leung}, \bibinfo{person}{Yunong Shi},
  \bibinfo{person}{David~I. Schuster}, \bibinfo{person}{Henry Hoffmann}, {and}
  \bibinfo{person}{Frederic~T. Chong}.} \bibinfo{year}{2019}\natexlab{}.
\newblock \showarticletitle{Partial Compilation of Variational Algorithms for
  Noisy Intermediate-Scale Quantum Machines}.
\newblock \bibinfo{journal}{\emph{Proceedings of the 52nd Annual IEEE/ACM
  International Symposium on Microarchitecture}} (\bibinfo{date}{Oct}
  \bibinfo{year}{2019}).
\newblock
\showISBNx{9781450369381}
\urldef\tempurl%
\url{https://doi.org/10.1145/3352460.3358313}
\showDOI{\tempurl}


\bibitem[Hahn(1950)]%
        {hahn}
\bibfield{author}{\bibinfo{person}{E.~L. Hahn}.}
  \bibinfo{year}{1950}\natexlab{}.
\newblock \showarticletitle{Spin Echoes}.
\newblock \bibinfo{journal}{\emph{Phys. Rev.}}  \bibinfo{volume}{80}
  (\bibinfo{date}{Nov} \bibinfo{year}{1950}), \bibinfo{pages}{580--594}.
\newblock
Issue 4.
\urldef\tempurl%
\url{https://doi.org/10.1103/PhysRev.80.580}
\showDOI{\tempurl}


\bibitem[Jurcevic et~al\mbox{.}(2021)]%
        {jurcevic2021demonstration}
\bibfield{author}{\bibinfo{person}{Petar Jurcevic}, \bibinfo{person}{Ali
  Javadi-Abhari}, \bibinfo{person}{Lev~S Bishop}, \bibinfo{person}{Isaac
  Lauer}, \bibinfo{person}{Daniela Borgorin}, \bibinfo{person}{Markus Brink},
  \bibinfo{person}{Lauren Capelluto}, \bibinfo{person}{Oktay Gunluk},
  \bibinfo{person}{Toshinari Itoko}, \bibinfo{person}{Naoki Kanazawa},
  {et~al\mbox{.}}} \bibinfo{year}{2021}\natexlab{}.
\newblock \showarticletitle{Demonstration of quantum volume 64 on a
  superconducting quantum computing system}.
\newblock \bibinfo{journal}{\emph{Quantum Science and Technology}}
  (\bibinfo{year}{2021}).
\newblock


\bibitem[Kandala et~al\mbox{.}(2017)]%
        {kandala2017hardware}
\bibfield{author}{\bibinfo{person}{Abhinav Kandala}, \bibinfo{person}{Antonio
  Mezzacapo}, \bibinfo{person}{Kristan Temme}, \bibinfo{person}{Maika Takita},
  \bibinfo{person}{Markus Brink}, \bibinfo{person}{Jerry~M Chow}, {and}
  \bibinfo{person}{Jay~M Gambetta}.} \bibinfo{year}{2017}\natexlab{}.
\newblock \showarticletitle{Hardware-efficient variational quantum eigensolver
  for small molecules and quantum magnets}.
\newblock \bibinfo{journal}{\emph{Nature}} \bibinfo{volume}{549},
  \bibinfo{number}{7671} (\bibinfo{year}{2017}), \bibinfo{pages}{242--246}.
\newblock


\bibitem[Khodjasteh and Lidar(2007)]%
        {DDKhodjasteh_2007}
\bibfield{author}{\bibinfo{person}{Kaveh Khodjasteh} {and}
  \bibinfo{person}{Daniel~A. Lidar}.} \bibinfo{year}{2007}\natexlab{}.
\newblock \showarticletitle{Performance of deterministic dynamical decoupling
  schemes: Concatenated and periodic pulse sequences}.
\newblock \bibinfo{journal}{\emph{Physical Review A}} \bibinfo{volume}{75},
  \bibinfo{number}{6} (\bibinfo{date}{Jun} \bibinfo{year}{2007}).
\newblock
\showISSN{1094-1622}
\urldef\tempurl%
\url{https://doi.org/10.1103/physreva.75.062310}
\showDOI{\tempurl}


\bibitem[Klimov et~al\mbox{.}(2018)]%
        {Klimov_2018}
\bibfield{author}{\bibinfo{person}{P.{\hspace{0.167em} }V. Klimov},
  \bibinfo{person}{J. Kelly}, \bibinfo{person}{Z. Chen}, \bibinfo{person}{M.
  Neeley}, \bibinfo{person}{A. Megrant}, \bibinfo{person}{B. Burkett},
  \bibinfo{person}{R. Barends}, \bibinfo{person}{K. Arya}, \bibinfo{person}{B.
  Chiaro}, \bibinfo{person}{Yu Chen}, \bibinfo{person}{A. Dunsworth},
  \bibinfo{person}{A. Fowler}, \bibinfo{person}{B. Foxen}, \bibinfo{person}{C.
  Gidney}, \bibinfo{person}{M. Giustina}, \bibinfo{person}{R. Graff},
  \bibinfo{person}{T. Huang}, \bibinfo{person}{E. Jeffrey},
  \bibinfo{person}{Erik Lucero}, \bibinfo{person}{J.{\hspace{0.167em}}Y.
  Mutus}, \bibinfo{person}{O. Naaman}, \bibinfo{person}{C. Neill},
  \bibinfo{person}{C. Quintana}, \bibinfo{person}{P. Roushan},
  \bibinfo{person}{Daniel Sank}, \bibinfo{person}{A. Vainsencher},
  \bibinfo{person}{J. Wenner}, \bibinfo{person}{T.{\hspace{0.167em}}C. White},
  \bibinfo{person}{S. Boixo}, \bibinfo{person}{R. Babbush},
  \bibinfo{person}{V.{\hspace{0.167em}}N. Smelyanskiy}, \bibinfo{person}{H.
  Neven}, {and} \bibinfo{person}{John{\hspace{0.167em}}M. Martinis}.}
  \bibinfo{year}{2018}\natexlab{}.
\newblock \showarticletitle{Fluctuations of Energy-Relaxation Times in
  Superconducting Qubits}.
\newblock \bibinfo{journal}{\emph{Physical Review Letters}}
  \bibinfo{volume}{121}, \bibinfo{number}{9} (\bibinfo{date}{aug}
  \bibinfo{year}{2018}).
\newblock
\urldef\tempurl%
\url{https://doi.org/10.1103/physrevlett.121.090502}
\showDOI{\tempurl}


\bibitem[LaRose et~al\mbox{.}(2021)]%
        {giurgicatiron2020digital}
\bibfield{author}{\bibinfo{person}{Ryan LaRose}, \bibinfo{person}{Andrea Mari},
  \bibinfo{person}{Sarah Kaiser}, \bibinfo{person}{Peter~J. Karalekas},
  \bibinfo{person}{Andre~A. Alves}, \bibinfo{person}{Piotr Czarnik},
  \bibinfo{person}{Mohamed~El Mandouh}, \bibinfo{person}{Max~H. Gordon},
  \bibinfo{person}{Yousef Hindy}, \bibinfo{person}{Aaron Robertson},
  \bibinfo{person}{Purva Thakre}, \bibinfo{person}{Nathan Shammah}, {and}
  \bibinfo{person}{William~J. Zeng}.} \bibinfo{year}{2021}\natexlab{}.
\newblock \bibinfo{title}{Mitiq: A software package for error mitigation on
  noisy quantum computers}.
\newblock
\newblock
\showeprint[arxiv]{2009.04417}~[quant-ph]


\bibitem[Lavrijsen et~al\mbox{.}(2020)]%
        {9259985}
\bibfield{author}{\bibinfo{person}{Wim Lavrijsen}, \bibinfo{person}{Ana Tudor},
  \bibinfo{person}{Juliane Müller}, \bibinfo{person}{Costin Iancu}, {and}
  \bibinfo{person}{Wibe de Jong}.} \bibinfo{year}{2020}\natexlab{}.
\newblock \showarticletitle{Classical Optimizers for Noisy Intermediate-Scale
  Quantum Devices}. In \bibinfo{booktitle}{\emph{2020 IEEE International
  Conference on Quantum Computing and Engineering (QCE)}}.
  \bibinfo{pages}{267--277}.
\newblock
\urldef\tempurl%
\url{https://doi.org/10.1109/QCE49297.2020.00041}
\showDOI{\tempurl}


\bibitem[Li and Benjamin(2017)]%
        {li2017efficient}
\bibfield{author}{\bibinfo{person}{Ying Li} {and} \bibinfo{person}{Simon~C.
  Benjamin}.} \bibinfo{year}{2017}\natexlab{}.
\newblock \showarticletitle{Efficient Variational Quantum Simulator
  Incorporating Active Error Minimization}.
\newblock \bibinfo{journal}{\emph{Phys. Rev. X}}  \bibinfo{volume}{7}
  (\bibinfo{date}{Jun} \bibinfo{year}{2017}), \bibinfo{pages}{021050}.
\newblock
Issue 2.
\urldef\tempurl%
\url{https://doi.org/10.1103/PhysRevX.7.021050}
\showDOI{\tempurl}


\bibitem[Lowe et~al\mbox{.}(2020)]%
        {zne4}
\bibfield{author}{\bibinfo{person}{Angus Lowe}, \bibinfo{person}{Max~Hunter
  Gordon}, \bibinfo{person}{Piotr Czarnik}, \bibinfo{person}{Andrew Arrasmith},
  \bibinfo{person}{Patrick~J Coles}, {and} \bibinfo{person}{Lukasz Cincio}.}
  \bibinfo{year}{2020}\natexlab{}.
\newblock \showarticletitle{Unified approach to data-driven quantum error
  mitigation}.
\newblock \bibinfo{journal}{\emph{arXiv preprint arXiv:2011.01157}}
  (\bibinfo{year}{2020}).
\newblock
\urldef\tempurl%
\url{https://arxiv.org/abs/2011.01157}
\showURL{%
\tempurl}


\bibitem[Martinis et~al\mbox{.}(2005)]%
        {martinis2005decoherence}
\bibfield{author}{\bibinfo{person}{John~M Martinis}, \bibinfo{person}{Ken~B
  Cooper}, \bibinfo{person}{Robert McDermott}, \bibinfo{person}{Matthias
  Steffen}, \bibinfo{person}{Markus Ansmann}, \bibinfo{person}{KD Osborn},
  \bibinfo{person}{Katarina Cicak}, \bibinfo{person}{Seongshik Oh},
  \bibinfo{person}{David~P Pappas}, \bibinfo{person}{Raymond~W Simmonds},
  {et~al\mbox{.}}} \bibinfo{year}{2005}\natexlab{}.
\newblock \showarticletitle{Decoherence in Josephson qubits from dielectric
  loss}.
\newblock \bibinfo{journal}{\emph{Physical review letters}}
  \bibinfo{volume}{95}, \bibinfo{number}{21} (\bibinfo{year}{2005}),
  \bibinfo{pages}{210503}.
\newblock


\bibitem[McClean et~al\mbox{.}(2016)]%
        {mcclean2016theory}
\bibfield{author}{\bibinfo{person}{Jarrod~R McClean}, \bibinfo{person}{Jonathan
  Romero}, \bibinfo{person}{Ryan Babbush}, {and} \bibinfo{person}{Al{\'a}n
  Aspuru-Guzik}.} \bibinfo{year}{2016}\natexlab{}.
\newblock \showarticletitle{The theory of variational hybrid quantum-classical
  algorithms}.
\newblock \bibinfo{journal}{\emph{New Journal of Physics}}
  \bibinfo{volume}{18}, \bibinfo{number}{2} (\bibinfo{year}{2016}),
  \bibinfo{pages}{023023}.
\newblock


\bibitem[Moll et~al\mbox{.}(2018)]%
        {moll2018quantum}
\bibfield{author}{\bibinfo{person}{Nikolaj Moll}, \bibinfo{person}{Panagiotis
  Barkoutsos}, \bibinfo{person}{Lev~S Bishop}, \bibinfo{person}{Jerry~M Chow},
  \bibinfo{person}{Andrew Cross}, \bibinfo{person}{Daniel~J Egger},
  \bibinfo{person}{Stefan Filipp}, \bibinfo{person}{Andreas Fuhrer},
  \bibinfo{person}{Jay~M Gambetta}, \bibinfo{person}{Marc Ganzhorn},
  {et~al\mbox{.}}} \bibinfo{year}{2018}\natexlab{}.
\newblock \showarticletitle{Quantum optimization using variational algorithms
  on near-term quantum devices}.
\newblock \bibinfo{journal}{\emph{Quantum Science and Technology}}
  \bibinfo{volume}{3}, \bibinfo{number}{3} (\bibinfo{year}{2018}),
  \bibinfo{pages}{030503}.
\newblock


\bibitem[M{\"u}ller et~al\mbox{.}(2019)]%
        {muller2019towards}
\bibfield{author}{\bibinfo{person}{Clemens M{\"u}ller},
  \bibinfo{person}{Jared~H Cole}, {and} \bibinfo{person}{J{\"u}rgen
  Lisenfeld}.} \bibinfo{year}{2019}\natexlab{}.
\newblock \showarticletitle{Towards understanding two-level-systems in
  amorphous solids: insights from quantum circuits}.
\newblock \bibinfo{journal}{\emph{Reports on Progress in Physics}}
  \bibinfo{volume}{82}, \bibinfo{number}{12} (\bibinfo{year}{2019}),
  \bibinfo{pages}{124501}.
\newblock


\bibitem[Murali et~al\mbox{.}(2019)]%
        {murali2019noise}
\bibfield{author}{\bibinfo{person}{Prakash Murali}, \bibinfo{person}{Jonathan~M
  Baker}, \bibinfo{person}{Ali Javadi-Abhari}, \bibinfo{person}{Frederic~T
  Chong}, {and} \bibinfo{person}{Margaret Martonosi}.}
  \bibinfo{year}{2019}\natexlab{}.
\newblock \showarticletitle{Noise-adaptive compiler mappings for noisy
  intermediate-scale quantum computers}. In
  \bibinfo{booktitle}{\emph{Proceedings of the Twenty-Fourth International
  Conference on Architectural Support for Programming Languages and Operating
  Systems}}. \bibinfo{pages}{1015--1029}.
\newblock


\bibitem[Murali et~al\mbox{.}(2020)]%
        {murali2020software}
\bibfield{author}{\bibinfo{person}{Prakash Murali}, \bibinfo{person}{David~C
  McKay}, \bibinfo{person}{Margaret Martonosi}, {and} \bibinfo{person}{Ali
  Javadi-Abhari}.} \bibinfo{year}{2020}\natexlab{}.
\newblock \showarticletitle{Software mitigation of crosstalk on noisy
  intermediate-scale quantum computers}. In
  \bibinfo{booktitle}{\emph{Proceedings of the Twenty-Fifth International
  Conference on Architectural Support for Programming Languages and Operating
  Systems}}. \bibinfo{pages}{1001--1016}.
\newblock


\bibitem[Peruzzo et~al\mbox{.}(2014)]%
        {peruzzo2014variational}
\bibfield{author}{\bibinfo{person}{Alberto Peruzzo}, \bibinfo{person}{Jarrod
  McClean}, \bibinfo{person}{Peter Shadbolt}, \bibinfo{person}{Man-Hong Yung},
  \bibinfo{person}{Xiao-Qi Zhou}, \bibinfo{person}{Peter~J Love},
  \bibinfo{person}{Al{\'a}n Aspuru-Guzik}, {and} \bibinfo{person}{Jeremy~L
  O’brien}.} \bibinfo{year}{2014}\natexlab{}.
\newblock \showarticletitle{A variational eigenvalue solver on a photonic
  quantum processor}.
\newblock \bibinfo{journal}{\emph{Nature communications}}  \bibinfo{volume}{5}
  (\bibinfo{year}{2014}), \bibinfo{pages}{4213}.
\newblock


\bibitem[Pokharel et~al\mbox{.}(2018a)]%
        {pokharel2018demonstration}
\bibfield{author}{\bibinfo{person}{Bibek Pokharel}, \bibinfo{person}{Namit
  Anand}, \bibinfo{person}{Benjamin Fortman}, {and} \bibinfo{person}{Daniel~A
  Lidar}.} \bibinfo{year}{2018}\natexlab{a}.
\newblock \showarticletitle{Demonstration of fidelity improvement using
  dynamical decoupling with superconducting qubits}.
\newblock \bibinfo{journal}{\emph{Physical review letters}}
  \bibinfo{volume}{121}, \bibinfo{number}{22} (\bibinfo{year}{2018}),
  \bibinfo{pages}{220502}.
\newblock


\bibitem[Pokharel et~al\mbox{.}(2018b)]%
        {DDPokharel_2018}
\bibfield{author}{\bibinfo{person}{Bibek Pokharel}, \bibinfo{person}{Namit
  Anand}, \bibinfo{person}{Benjamin Fortman}, {and} \bibinfo{person}{Daniel~A.
  Lidar}.} \bibinfo{year}{2018}\natexlab{b}.
\newblock \showarticletitle{Demonstration of Fidelity Improvement Using
  Dynamical Decoupling with Superconducting Qubits}.
\newblock \bibinfo{journal}{\emph{Physical Review Letters}}
  \bibinfo{volume}{121}, \bibinfo{number}{22} (\bibinfo{date}{Nov}
  \bibinfo{year}{2018}).
\newblock
\showISSN{1079-7114}
\urldef\tempurl%
\url{https://doi.org/10.1103/physrevlett.121.220502}
\showDOI{\tempurl}


\bibitem[Preskill(2018)]%
        {preskill2018quantum}
\bibfield{author}{\bibinfo{person}{John Preskill}.}
  \bibinfo{year}{2018}\natexlab{}.
\newblock \showarticletitle{Quantum Computing in the NISQ era and beyond}.
\newblock \bibinfo{journal}{\emph{Quantum}}  \bibinfo{volume}{2}
  (\bibinfo{year}{2018}), \bibinfo{pages}{79}.
\newblock


\bibitem[Ravi et~al\mbox{.}(2021)]%
        {ravi2021vaqem}
\bibfield{author}{\bibinfo{person}{Gokul~Subramanian Ravi},
  \bibinfo{person}{Kaitlin~N. Smith}, \bibinfo{person}{Pranav Gokhale},
  \bibinfo{person}{Andrea Mari}, \bibinfo{person}{Nathan Earnest},
  \bibinfo{person}{Ali Javadi-Abhari}, {and} \bibinfo{person}{Frederic~T.
  Chong}.} \bibinfo{year}{2021}\natexlab{}.
\newblock \bibinfo{title}{VAQEM: A Variational Approach to Quantum Error
  Mitigation}.
\newblock
\newblock
\showeprint[arxiv]{2112.05821}~[quant-ph]


\bibitem[Robey et~al\mbox{.}(1992)]%
        {CFAR}
\bibfield{author}{\bibinfo{person}{F.C. Robey}, \bibinfo{person}{D.R.
  Fuhrmann}, \bibinfo{person}{E.J. Kelly}, {and} \bibinfo{person}{R.
  Nitzberg}.} \bibinfo{year}{1992}\natexlab{}.
\newblock \showarticletitle{A CFAR adaptive matched filter detector}.
\newblock \bibinfo{journal}{\emph{IEEE Trans. Aerospace Electron. Systems}}
  \bibinfo{volume}{28}, \bibinfo{number}{1} (\bibinfo{year}{1992}),
  \bibinfo{pages}{208--216}.
\newblock
\urldef\tempurl%
\url{https://doi.org/10.1109/7.135446}
\showDOI{\tempurl}


\bibitem[Schl{\"o}r et~al\mbox{.}(2019)]%
        {schlor2019correlating}
\bibfield{author}{\bibinfo{person}{Steffen Schl{\"o}r},
  \bibinfo{person}{J{\"u}rgen Lisenfeld}, \bibinfo{person}{Clemens M{\"u}ller},
  \bibinfo{person}{Alexander Bilmes}, \bibinfo{person}{Andre Schneider},
  \bibinfo{person}{David~P Pappas}, \bibinfo{person}{Alexey~V Ustinov}, {and}
  \bibinfo{person}{Martin Weides}.} \bibinfo{year}{2019}\natexlab{}.
\newblock \showarticletitle{Correlating decoherence in transmon qubits: Low
  frequency noise by single fluctuators}.
\newblock \bibinfo{journal}{\emph{Physical review letters}}
  \bibinfo{volume}{123}, \bibinfo{number}{19} (\bibinfo{year}{2019}),
  \bibinfo{pages}{190502}.
\newblock


\bibitem[Smith et~al\mbox{.}(2021)]%
        {smith2021error}
\bibfield{author}{\bibinfo{person}{Kaitlin~N Smith},
  \bibinfo{person}{Gokul~Subramanian Ravi}, \bibinfo{person}{Prakash Murali},
  \bibinfo{person}{Jonathan~M Baker}, \bibinfo{person}{Nathan Earnest},
  \bibinfo{person}{Ali Javadi-Abhari}, {and} \bibinfo{person}{Frederic~T
  Chong}.} \bibinfo{year}{2021}\natexlab{}.
\newblock \showarticletitle{Error Mitigation in Quantum Computers through
  Instruction Scheduling}.
\newblock \bibinfo{journal}{\emph{arXiv preprint arXiv:2105.01760}}
  (\bibinfo{year}{2021}).
\newblock


\bibitem[Souza et~al\mbox{.}(2012)]%
        {souza2012robust}
\bibfield{author}{\bibinfo{person}{Alexandre~M Souza},
  \bibinfo{person}{Gonzalo~A {\'A}lvarez}, {and} \bibinfo{person}{Dieter
  Suter}.} \bibinfo{year}{2012}\natexlab{}.
\newblock \showarticletitle{Robust dynamical decoupling}.
\newblock \bibinfo{journal}{\emph{Philosophical Transactions of the Royal
  Society A: Mathematical, Physical and Engineering Sciences}}
  \bibinfo{volume}{370}, \bibinfo{number}{1976} (\bibinfo{year}{2012}),
  \bibinfo{pages}{4748--4769}.
\newblock


\bibitem[Tannu and Qureshi(2019a)]%
        {tannu2019mitigating}
\bibfield{author}{\bibinfo{person}{Swamit~S Tannu} {and}
  \bibinfo{person}{Moinuddin~K Qureshi}.} \bibinfo{year}{2019}\natexlab{a}.
\newblock \showarticletitle{Mitigating measurement errors in quantum computers
  by exploiting state-dependent bias}. In \bibinfo{booktitle}{\emph{Proceedings
  of the 52nd Annual IEEE/ACM International Symposium on Microarchitecture}}.
  \bibinfo{pages}{279--290}.
\newblock


\bibitem[Tannu and Qureshi(2019b)]%
        {tannu2019not}
\bibfield{author}{\bibinfo{person}{Swamit~S Tannu} {and}
  \bibinfo{person}{Moinuddin~K Qureshi}.} \bibinfo{year}{2019}\natexlab{b}.
\newblock \showarticletitle{Not all qubits are created equal: a case for
  variability-aware policies for NISQ-era quantum computers}. In
  \bibinfo{booktitle}{\emph{Proceedings of the Twenty-Fourth International
  Conference on Architectural Support for Programming Languages and Operating
  Systems}}. \bibinfo{pages}{987--999}.
\newblock


\bibitem[Temme et~al\mbox{.}(2017)]%
        {temme2017error}
\bibfield{author}{\bibinfo{person}{Kristan Temme}, \bibinfo{person}{Sergey
  Bravyi}, {and} \bibinfo{person}{Jay~M Gambetta}.}
  \bibinfo{year}{2017}\natexlab{}.
\newblock \showarticletitle{Error mitigation for short-depth quantum circuits}.
\newblock \bibinfo{journal}{\emph{Physical review letters}}
  \bibinfo{volume}{119}, \bibinfo{number}{18} (\bibinfo{year}{2017}),
  \bibinfo{pages}{180509}.
\newblock


\bibitem[Uvarov et~al\mbox{.}(2020)]%
        {uvarov2020machine}
\bibfield{author}{\bibinfo{person}{AV Uvarov}, \bibinfo{person}{AS Kardashin},
  {and} \bibinfo{person}{Jacob~D Biamonte}.} \bibinfo{year}{2020}\natexlab{}.
\newblock \showarticletitle{Machine learning phase transitions with a quantum
  processor}.
\newblock \bibinfo{journal}{\emph{Physical Review A}} \bibinfo{volume}{102},
  \bibinfo{number}{1} (\bibinfo{year}{2020}), \bibinfo{pages}{012415}.
\newblock


\bibitem[Vancoillie et~al\mbox{.}(2016)]%
        {vancoillie2016potential}
\bibfield{author}{\bibinfo{person}{Steven Vancoillie},
  \bibinfo{person}{Per~{\AA}ke Malmqvist}, {and} \bibinfo{person}{Valera
  Veryazov}.} \bibinfo{year}{2016}\natexlab{}.
\newblock \showarticletitle{Potential energy surface of the chromium dimer
  re-re-revisited with multiconfigurational perturbation theory}.
\newblock \bibinfo{journal}{\emph{Journal of chemical theory and computation}}
  \bibinfo{volume}{12}, \bibinfo{number}{4} (\bibinfo{year}{2016}),
  \bibinfo{pages}{1647--1655}.
\newblock


\bibitem[Viola et~al\mbox{.}(1999)]%
        {viola1999dynamical}
\bibfield{author}{\bibinfo{person}{Lorenza Viola}, \bibinfo{person}{Emanuel
  Knill}, {and} \bibinfo{person}{Seth Lloyd}.} \bibinfo{year}{1999}\natexlab{}.
\newblock \showarticletitle{Dynamical decoupling of open quantum systems}.
\newblock \bibinfo{journal}{\emph{Physical Review Letters}}
  \bibinfo{volume}{82}, \bibinfo{number}{12} (\bibinfo{year}{1999}),
  \bibinfo{pages}{2417}.
\newblock


\bibitem[Welch et~al\mbox{.}(1995)]%
        {kalman}
\bibfield{author}{\bibinfo{person}{Greg Welch}, \bibinfo{person}{Gary Bishop},
  {et~al\mbox{.}}} \bibinfo{year}{1995}\natexlab{}.
\newblock \showarticletitle{An introduction to the Kalman filter}.
\newblock  (\bibinfo{year}{1995}).
\newblock


\end{thebibliography}

\end{document}